

\def\footnoterule{\kern -3pt \hrule width 0truein \kern 2.4pt}
\newcount\notenumber \notenumber=1
\def\note#1{\footnote{$^{\the\notenumber}$}{#1}\global\advance\notenumber by 1}

\font\forteenrmb=cmbx10 at 14pt
\font\twelvermb=cmbx10 at 12pt
\font\elevenrmb=cmbx10 at 11pt
\font\eightrmb=cmbx10 at 8pt

\bigskip
\bigskip
\noindent{\forteenrmb Block orthogonal polynomials: I.}

\noindent{\twelvermb Definitions and properties}

\bigskip
\bigskip
{\leftskip=1.0cm \noindent {\bf Jean-Marie Normand}

\medskip
{\eightrmb \noindent Service de Physique Th\'eorique, CEA/DSM/SPhT -- CNRS/MIPPU/URA 2306

\noindent CEA/Saclay, F-91191 Gif-sur-Yvette Cedex, France

\medskip

\noindent E-mail: jean-marie.normand@cea.fr
}

\bigskip
\bigskip
\noindent{\bf Abstract}

\noindent Constrained orthogonal polynomials have been recently introduced in the study of the
Hohenberg-Kohn functional to provide basis functions satisfying particle number conservation
for an expansion of the particle density.
More generally, we define block orthogonal (BO) polynomials which are orthogonal, with respect to a first
Euclidean inner product, to a given $i$-dimensional subspace ${\cal E}_i$ of polynomials associated with
the constraints.
In addition, they are mutually orthogonal with respect to a second Euclidean inner
product.
We recast the determination of these polynomials into a general problem of finding particular orthogonal
bases in an Euclidean vector space endowed with distinct inner products.
An explicit two step Gram-Schmidt orthogonalization (G-SO) process to determine these bases is given.
By definition, the standard block orthogonal (SBO) polynomials are associated with a choice of ${\cal E}_i$
equal to the subspace of polynomials of degree less than $i$.
We investigate their properties, emphasizing similarities to and differences from the standard orthogonal
polynomials.
Applications to classical orthogonal polynomials will be given in forthcoming papers.

\par}

\bigskip
\bigskip
\noindent{\elevenrmb 1. Introduction}

\bigskip
\noindent Recently, B Giraud {\it et al.} [1--3] have considered new sets of
{\it constrained orthogonal polynomials}.
Basically, these real polynomials $P_n(x),\,n=1,2,\ldots$ of {\it exact} degree $n$
(i.e. the coefficient of $x^n$ is nonzero), satisfy the constraint of vanishing average
\note{A constraint of `vanishing momentum', $\int_a^b dx\,w(x) x P_n(x)=0$,
is also considered in [1] section 3.
See footnote 44.}
$\,$with a non-negative weight function $w(x)$ on a real interval $[a,b]$,
$$\int_a^b dx\,w(x) P_n(x)=0\quad n=1,2,\ldots\,. \eqno(1.1)$$
In addition, they are orthogonal on the same interval with a distinct non-negative weight function $w_2(x)$,
$$\int_a^b dx\,w_2(x) P_m(x) P_n(x)\propto\delta_{m,n}\quad m,n=1,2,\ldots\,. \eqno(1.2)$$
The constant polynomial $P_0$ is excluded since it does not fit the constraint.
As a result, these polynomials does not form a complete set, but span a subspace which can be well-suited
to specific applications.
Thus, in various problems of mathematical physics, one considers an unknown function $f(x)$ which
has to satisfy a similar constraint, $\int_a^b dx\,w(x) f(x)=0$.
One way to take into account this constraint readily, is to expand $f(x)$ in terms of the basis
polynomials $P_n(x),\,n=1,2,\ldots$, as $f(x)=\sum_{n=1}^{n_c} f_n P_n(x)$, with $n_c$ a possible cut-off.
The choice of the weight functions is dictated by the problem investigated and physical considerations,
giving, possibly, a physical meaning both to the polynomials and to the possible truncation of
the functional space.

This method has been applied to the {\it Hohenberg-Kohn variational principle} [4]
for the ground stground stateate energy, originally established for an interacting electron gas in a local spin-independent
external potential $v({\bf r})$ leading to a non-degenerate ground state.
The variable function is the particle density $n({\bf r})$.
The Hamiltonian reads ${\cal H}=T+V+U$ where $T$ is the kinetic energy, $V$ corresponds
to the local one-particle external potential $v({\bf r})$ and $U$ is the two-particle interaction
(e.g., the Coulomb interaction).
With number of particles $N$, and their mutual interaction $U$ specified,
let ${\cal V}$ be a set of local one-particle potentials $v({\bf r})$ such that, for each $v({\bf r})$,
there exists a non-degenerate $N$-particle ground state $|\Psi\rangle$ with energy $E$,
that is solution of the Schr\"odinger equation ${\cal H}|\Psi\rangle=E|\Psi\rangle$.
From $|\Psi\rangle$, one calculates the ground state particle density $n({\bf r})$,
see, e.g., equation (1.11) below.
By definition, the number of particles is a functional of $n({\bf r})$ such that
$$N[n]:=\int d{\bf r}\,n({\bf r})=N\,. \eqno(1.3)$$
Thereby, one defines the mappings $v({\bf r})\mapsto|\Psi\rangle\mapsto n({\bf r})$.
Let the set ${\cal N}$ of particle density functions be defined as the image of ${\cal V}$ in the resulting mapping
$v({\bf r})\mapsto n({\bf r})$.
Since a real additive constant in $v({\bf r})$ has no effect on the ground state $|\Psi\rangle$ and thus on
$n({\bf r})$, potentials in ${\cal V}$ differing by a real constant are considered equivalent.
Let $\{v({\bf r})\}$ denotes the equivalence class.
Then, the {\it Hohenberg-Kohn theorem} states that
{\sl

\noindent (i) for all $v({\bf r})$ in ${\cal V}$ and all $n({\bf r})$ in ${\cal N}$, there are one-to-one
correspondences $\{v({\bf r})\}\,\leftrightarrow\,|\Psi\rangle\,\leftrightarrow\,n({\bf r})$;

\noindent (ii) according to (i), let $|\Psi\rangle$ be the $N$-particle ground state corresponding to any
$n({\bf r})$ in ${\cal N}$, then,
$$\forall\,n({\bf r})\in{\cal N},\,n({\bf r})\,\mapsto\,|\Psi\rangle
\qquad F[n]:=\langle\Psi|T+U|\Psi\rangle \eqno(1.4)$$
is a universal functional, i.e. it does not depend on the external potential $v({\bf r})$
and is valid for any number of particles $N=N[n]$.
Now, for any $v_0({\bf r})$ in ${\cal V}$, the energy functional,
$${\cal E}_{v_0}[n]:=\int d{\bf r}\,v_0({\bf r}) n({\bf r})+F[n] \eqno(1.5)$$
assumes its minimum value on ${\cal N}$ for the exact ground state density $n_0({\bf r})$ corresponding to
$v_0({\bf r})$ according to (i)}
\note{The proof relies on the Rayleigh-Ritz minimum principle for the energy
functional $E[\Psi]:=\langle\Psi|H|\Psi\rangle$, in which the wavefunction $\Psi$ is the variable function.}
{\sl.
Then, ${\cal E}_{v_0}[n_0]$ is equal to the corresponding N-particle ground state energy
$E_0=\min_{n\in{\cal N}}{\cal E}_{v_0}[n]$.
}

\noindent Let us note that the implementation of this variational principle raises several fundamental problems
about:
(i) the possible extension of ${\cal N}$ to wider spaces of trial particle density;
(ii) the existence of a functional derivative to formulate the variational principle as,
 $\delta {\cal E}_{v_0}[n]/\delta n({\bf r})=0$ [5]
\note{See, e.g., [5] sections 2.1 and 2.3 and the references therein, especially the works of Lieb and Levy.
The particle density $n({\bf r})\in{\cal N}$ considered in the Hohenberg-Kohn theorem is said to be a
{\it pure state v-representable} since it is the density of a pure ground state $|\Psi\rangle$ for a specific
local external potential $v({\bf r})\in{\cal V}$.
Now, for any given non-negative $n({\bf r})$, normalized according to equation (1.3), does there exists a
local external potential $v({\bf r})$ so that $v({\bf r})\mapsto n({\bf r})$?
The answer can be negative, even when the formalism is extended to non-pure N-particle states described by
density operators.}.

Based on the work of Hohenberg and Kohn  and several extensions [5]
\note{Degenerate ground state, finite temperature ensembles, etc.; see, e.g., [5] section 3 and
the references listed in this review book.},
the {\it density functional theory} has become a standard approach to investigate the properties of quantum
interacting many-particle systems in terms of the particle density $n({\bf r})$.
This method has been frequently used in several branches of chemistry and physics, e.g., in atomic, molecular
and nuclear theoretical physics.
Along the Thomas-Fermi approach and the work of Kohn and Sham [6] (replacing direct variations
with respect to the particle density by an auxiliary orbital picture), several functionals
have been successfully used [5, 7].
Then, a constructive study of the ground state is provided by standard perturbation theories
(e.g., particle-hole excitations, configuration mixing), around a mean field first order
approximation.
The success of these methods is based on the existence of suitable truncations to relevant subspaces.

In view of their use in some problems of nuclear physics, B Giraud {\it et al} aim to make a similar approach
in the space of particle density functions instead of the space of wavefunctions.
In other words, can the functional space of $n({\bf r})$ be truncated to a subset of meaningful
`density modes'? [1--3].
An ultimate goal would be a constructive approach to the Hohenberg-Kohn functional.
As a first step, the one-to-one correspondence between the particle density and the external potential
(up to within an additive constant)
is investigated through the related values of $\delta n({\bf r})$ and $\delta v({\bf r})$
around the true solution $n_0({\bf r})$ for a given $v_0({\bf r})$.
Then, the question is: can the functional spaces for $\delta n({\bf r})$ and for $\delta v({\bf r})$ be defined by
`suitable' basis functions, such that the truncation of theses functional spaces to trial subspaces of few basis
functions (or `modes') be relevant?
In heuristic approach, forgetting the questions brought up previously about the functional spaces,
we focus on the following basic constraints:
(i) $n({\bf r})$ has to be non-negative and the particle number conservation (1.3) yields
$$\int d{\bf r}\,\delta n({\bf r})=0\,; \eqno(1.6)$$
(ii) the one-to-one mapping is between $\{v({\bf r})\}$ and $n({\bf r})$, therefore $\delta v({\bf r})$
must not be a nonzero constant,
$$\delta v({\bf r})\ne{\rm const}\,. \eqno(1.7)$$
A standard way to take into account the particle number conservation with the variational formulation is to
introduce a Lagrange multiplier $\mu$ such that
$\delta({\cal E}_{v_0}[n]-\mu \int d{\bf r}\,n({\bf r}))/\delta n({\bf r})=0$.
Once again, this requires an extension of the functional space to particle density normalized
to a not-necessarily-integer value as in equation (1.3)
\note{See, e.g., [5] section 2.4 and [7] section 1.4.5, the problem of derivative discontinuities.}.
The method sketched in our first paragraph presents an alternative overcoming these difficulties
by considering trial variations $\delta n({\bf r})$ which inherently satisfy the constraint (1.6).
This is done as follows in a simple toy model.
As a usual first approximation in the nuclear shell model,
a system of $N$ independent fermions in a one-dimensional harmonic oscillator potential $v_0(x):={1\over2}\,x^2$
is considered in [1--3]
\note{See [1] section 4, [2] section 4 and [3] section 2.},
neglecting the two-body interactions.
Then, the single-particle orbital wavefunctions are defined by [8]
\note{See, e.g., [8] appendix B (B.65) and (B.70).},
$$\eqalignno{
h &:={1\over2} \Bigl(-{d^2\over dx^2}+x^2\Bigr)\qquad h\,\psi_j(x)=\epsilon_j\,\psi_j(x) &(1.8)\cr
\psi_j(x) &\phantom{:}=(2^n n!)^{-{1\over2}}\,\pi^{-{1\over4}}\,e^{-{1\over2}x^2} H_j(x)
\qquad\epsilon_j={1\over2}+j\quad j=0,1,\ldots &(1.9)\cr
}$$
where $H_j(x)$ is the Hermite polynomial.
The $N$-particle ground state wavefunction $\Psi_0(x_1,\ldots,x_N)$ is the properly normalized $N\times N$
Slater determinant built from the $N$ lowest energy single-particle eigenfunctions,
$$\Psi_0(x_1,\ldots,x_N)
=(N!)^{-{1\over2}} \det\bigl[\psi_j(x_k)\bigr]_{j=0,\ldots,N-1\atop k=1,\ldots,N\hfill}\,. \eqno(1.10)$$
The particle density, satisfying the normalization (1.3), reads
$$n_0(x):=N \int_{-\infty}^\infty dx_2\cdots\int_{-\infty}^\infty dx_N |\Psi_0(x,x_2,\ldots,x_N)|^2
=\sum_{j=0}^{N-1} \psi_j(x)^2\,. \eqno(1.11)$$
Now, it follows from the first order perturbation theory [8]
\note{See, e.g., [8] chapter XVI, section 2 (XVI.16).}
$\,$that the response to a variation $\delta v(x)$ is for each single-particle wavefunction,
$$\delta\psi_j(x)=\sum_{J=N}^\infty \psi_J(x) {\langle\psi_J|\delta v|\psi_j\rangle\over\epsilon_j-\epsilon_J}
\qquad\langle\psi_J|\delta v|\psi_j\rangle=\int_{-\infty}^\infty dx\,\psi_J(x) \delta v(x) \psi_j(x)
\eqno(1.12)$$
where $j=0,\ldots,N-1$ and $J=N,\ldots,\infty$ are the hole and particle indices, respectively.
As a result, from equation (1.11), the variation of the particle density is
\note{The positivity of the density $\rho_0+\delta\rho$ is guaranteed at first order, since $\delta\rho$ follows from a
first order calculation of the perturbed N-particle wavefunction $\Psi_0+\delta\Psi$.}
$$\delta n(x)=2\,\sum_{j=0}^{N-1} \sum_{J=N}^\infty \psi_j(x) \psi_J(x)
{\langle\psi_J|\delta v|\psi_j\rangle\over\epsilon_j-\epsilon_J} \eqno(1.13)$$
thereby, providing the functional mapping $\delta v(x)\mapsto\delta n(x)$, linear in first order
perturbation theory.
As usual, for practical study, this functional correspondence can be transformed into a discrete
(possibly infinite) linear problem
by expanding both $\delta n(x)=\sum_{j=0}^\infty \delta n_j \varphi_j(x)$ and
$\delta v(x)=\sum_{k=0}^\infty \delta v_k \phi_k(x)$ on suitable bases, $\{\varphi_n(x),\,n=0,1,\ldots\}$ and
$\{\phi_n(x),\,n=0,1,\ldots\}$, respectively.
Now, since $\psi_j(x) \psi_J(x)\propto\exp(-x^2)$ times a polynomial, $\delta\rho(x)$ can be expanded readily
on functions $\varphi_n(x):=\exp(-x^2) P_n(x)$, with $\{P_n(x),\,n=0,1,\ldots\}$ any basis of real polynomials.
It is {\it convenient} to choose these polynomials of degree $n$ and such that the functions
$\varphi_n(x),\,n=0,1,\ldots$ be orthogonal,
$$\int_{-\infty}^\infty dx\,\varphi_m(x) \varphi_n(x)
=\int_{-\infty}^\infty dx\,e^{-2x^2} P_m(x) P_n(x)\propto\delta_{m,n}\,. \eqno(1.14)$$
This amounts to choose the polynomials $P_n(x),\,n=0,1,\ldots$ to be orthogonal with respect to
the weight function $w_2(x):=\exp(-2x^2)$.
In addition, the particle number conservation (1.6) will be satisfied trivially if each basis function
$\varphi_n(x)$ fulfils the constraint,
$$\int_{-\infty}^\infty dx\,\varphi_n(x)=\int_{-\infty}^\infty dx\,e^{-x^2}\,P_n(x)
\propto\int_{-\infty}^\infty dx\,e^{-x^2}\,P_0 P_n(x)=0 \eqno(1.15)$$
i.e., if the polynomial $P_n(x)$ is orthogonal to the constant polynomial $P_0$ with
the weight function $w(x):=\exp(-x^2)$.
As already noted, this excludes the value $n=0$.
Thus, for this toy model within the framework of perturbation theory and particle-hole excitations,
{\it the functions $\varphi_n(x)=w(x) P_n(x)$, with polynomials $P_n(x),\,n=1,2,\ldots$ satisfying
equations (1.1) and (1.2),
assuming such polynomials exist, provide a functional space for $\delta n(x)$ taking into account the
particle number conservation}, essential for the Hohenberg-Kohn variational principle.
Now, what about the choice of the basis functions $\phi_n(x)$ for the expansion of $\delta v(x)$?
Although the constraints (1.6) and (1.7) are different, it is {\it convenient} to choose the same basis, i.e.
$\phi_n(x):=\varphi_n(x),\,n=1,2,\ldots$ (otherwise, connection coefficients between the two bases would be
required in the calculation).
Nevertheless, it should be noted that the constraint (1.6), $\int_{-\infty}^\infty dx\,\delta v(x)=0$,
implies $\delta v(x)$ satisfies the constraint (1.7), but this not a necessary condition
\note{B Giraud {\it et al} argue in [1] section 1, [2] section 1 and [3] section 2,
that `the basis must be orthogonal to a flat potential' (i.e. constant).
This is not true, e.g., $\delta v(x)=\varphi_0(x)\propto w(x)$ does not fit the constraint (1.6), but satisfies
the constraint (1.7).
A basis excluding a constant function could be $\{x^n,\,n=1,2,\ldots\}$.},
and therefore, this corresponds to a particular restriction of the functional space for $\delta v(x)$.
Then, it is argued in [1--3] that these basis functions are good candidates to define relevant
particle density modes.
The same approach can be generalized for any one-particle potential $v_0({\bf r})$,
not only in a one-dimensional model.
Thus, in connection with the Laguerre polynomials, the following basis functions and weight functions
are considered in [2]
\note{For example, in atomic and molecular physics, one may think about the Coulomb interaction
$v_0({\bf r})\propto 1/x$, where $x$ is the radial variable in a $3$-dimensional space.
The single-particle orbital wavefunction includes an exponential term linear in $x$,
see, e.g., [8] chapter XI, section 6.
Then, it would be clumsy to consider a basis function $\varphi_n(x)$ with an exponential term quadratic in $x$,
as for the harmonic oscillator.},
$$\varphi_n(x):=e^{-x} P_n(x)\qquad w(x):=e^{-x} x^{d-1}\qquad w_2(x):=e^{-2x} x^{d-1}\,. \eqno(1.16)$$
The functions $\varphi_n(x),\,n=1,2,\ldots$ have to satisfy the orthogonality relations (1.14) and (1.15) with now
$x\in[0,\infty)$ as, e.g., the radial variable in a $d$-dimensional space.

The purpose of this paper is to make a systematic study of this kind of polynomials defined by equations
(1.1) and (1.2).
The problem can be generalized as follows.
For given nonzero and non-negative weight functions $w(x)$ and $w_2(x)$ on a given interval,
and for given $i$ linearly independent polynomials $e_1(x),\ldots,e_i(x)$
(possibly associated with $i$ `constraints'
\note{For equation (1.1), $i:=1$ and $e_1(x):=P_0={\rm const}$.}), spanning a subspace ${\cal E}_1$
of a space ${\cal E}$ of polynomials:
(i) does there exist a subspace ${\cal E}_2$ orthogonal to ${\cal E}_1$ with respect to the weight function $w(x)$,
together with ${\cal E}_1$ and ${\cal E}_2$ complementary in ${\cal E}$?
(ii) is it possible to define, and compute, within ${\cal E}_2$ an orthogonal basis with
respect to the other weight function $w_2(x)$?
Thereby, one defines what we call {\it block orthogonal} (BO) {\it polynomials},
instead of constrained orthogonal polynomials,
to underlined the fact that the linear constraints are defined by an inner product.
Although the problem was motivated by polynomial considerations, it is worthwhile to point out
what can be generalized and ascribed to Euclidean vector space and to polynomial algebra, respectively.

This paper is organized as follows.
The general problem in linear algebra is considered in section 2.
For the sake of clarity, elementary results for Euclidean spaces are recalled in section 2.1
(see, e.g. [9--11]).
Two BO subspaces and bases are defined and constructed in section 2.2.
The well-defined method is based on a two step Gram-Schmidt orthogonalization (G-SO) process.
The possible extension to more than two subspaces is discussed in section 2.3.
These formal algebraic considerations are applied to polynomial vector spaces in section 3.
After generalities about Euclidean vector spaces of polynomials in section 3.1 (see, e.g. [12--14]),
BO polynomials for two subspaces are constructed in section 3.2.
Then, we focus, for a given non-negative integer $i$, on the real BO polynomials
$P_{i;n}(x),\,n=i,i+1,\ldots$ of exact degree $n$, orthogonal to the polynomials of degree less than $i$
with the weight function $w(x)$, and mutually orthogonal with respect to the weight function $w_2(x)$
\note{In other words, the $i$ linear constraints are associated with the $i$ linearly independent polynomials
$e_{j-1}(x):=x^j,\,j=0,\ldots,i-1$, defining the subspace ${\cal P}_i:={\cal E}_1$ of polynomials
of degree less than $i$.},
$$\eqalignno{
\int_a^b dx\,w(x) x^m P_{i;n}(x) &=0\quad m=0,\ldots,i-1\quad n=i,i+1,\ldots &(1.17)\cr
\int_a^b dx\,w_2(x) P_{i;m}(x) P_{i;n}(x) &\propto\delta_{m,n}\quad m,n=i,i+1,\ldots &(1.18)\cr
}$$
generalizing equations (1.1) and (1.2) for $i\ge1$.
We call these polynomials {\it standard block orthogonal} (SBO) {\it polynomials}.
Their general properties are investigated in section 3.3, comparing them with the properties
of the {\it standard orthogonal polynomials}.
We give our conclusions in section 4.
For completeness, and to underline the similarities to and the differences from the study of BO
polynomials, definition and properties of standard orthogonal polynomials are recalled in appendix A.
Examples of BO polynomials for three subspaces are given in appendix B.

Throughout the remaining of this paper, the following conventions and notations are used:

\noindent -- $i,j,k,\ell,m,n$ and $N$ denote non-negative integers;

\noindent -- a null sum is interpreted as zero while a null product or a null determinant
is interpreted as unity;

\noindent -- matrices are denoted in boldface, e.g., ${\bf A}:=(A_{j,k})_{j,k=\cdots}$;

\noindent -- for any function $f:x\mapsto f(x)$ (especially a polynomial), $f$ stands for $f(x)$;

\noindent -- {\it monic polynomials}, i.e. with the coefficient of the highest power equal to one,
and also any related quantities are denoted by hatted letters, e.g., $\widehat P_n$
and $\widehat h_n$;

\noindent -- in the differentiation and/or recurrence formulae, the polynomials with negative
degree are set equal to zero, e.g., $P_{-2}=P_{-1}:=0$;

\noindent -- standard orthogonal polynomials are denoted by $Q_n,\,n=0,1,\ldots$, where
$Q_n$ is of exact degree $n$;

\noindent -- {\it classical orthogonal polynomials} (e.g., the Hermite and Laguerre polynomials)
are defined according to [14--16]
\note{See, [14] chapter X, [15] chapter 22 or [16] section 8.9.};

\noindent -- BO polynomials are denoted with a capital $P$, e.g., $P_{i;n}$ for SBO polynomials.

\medskip
\bigskip
\noindent{\elevenrmb 2. Block orthogonal subspaces and bases}

\bigskip
\noindent{\it 2.1. Basic definitions and properties of complementary and orthogonal subspaces}

\medskip
\noindent Let ${\cal E}$ be an $N$-dimensional vector space over the real field $R$.
A symmetric and positive-definite (thus non-degenerate)
bilinear form on ${\cal E}$, defines an {\it Euclidean inner product} or {\it scalar product} $(\,,\,)$ such that
\note{The bilinear form is said to be:
(i) {\it non-degenerate} if for every $u$, $(u\,,\,v)=0$ implies $v=0$;
(ii) {\it definite} if $(u\,,\,u)=0$ implies $u=0$;
(iii) {\it positive} if $(u\,,\,u)$ is positive for every nonzero $u$.
See, e.g., [9] sections 59--61, [10] chapter XIV, section 1 or [11] 3.1-1.},
$$\forall\,u,v\in {\cal E}\quad(u\,,\,v)=(v\,,\,u)\in R\qquad
(u\,,\,u)\ge0\quad{\rm and}\quad(u\,,\,u)=0\ \Leftrightarrow\ u=0\,. \eqno(2.1)$$
Any $N_1$-dimensional subspace ${\cal E}_1$ of ${\cal E}$, not $\{0\}$ or ${\cal E}$ itself,
has an infinity of {\it complementary subspaces}, all of dimension $N_2={\rm codim}_{\cal E}\,{\cal E}_1:=N-N_1$,
i.e. the {\it codimension}
\note{See, e.g., [17] A, II, section 7, no 3, definition 2, p 99.}
of ${\cal E}_1$ with respect to ${\cal E}$.
Indeed, any basis $\{e_1,\ldots,e_{N_1}\}$ of ${\cal E}_1$ can be completed with $N_2$ vectors
$\varepsilon_1,\ldots,\varepsilon_{N_2}$ such that $\{e_1,\ldots,e_{_1},\varepsilon_1,\ldots,\varepsilon_{N_2}\}$
is a basis of ${\cal E}$.
Then, $\varepsilon_1,\ldots,\varepsilon_{N_2}$ generate a complementary subspace ${\cal E}_2$ of ${\cal E}_1$.
For any arbitrary vectors $u_1,\ldots,u_{N_2}$ in ${\cal E}_1$, not all zero, the $N_2$ vectors,
$\varepsilon'_j:=u_j+\varepsilon_j,\,j=1,\ldots,N_2$ generate another complementary subspace ${\cal E}'_2$
of  ${\cal E}_1$.
Now, it is known that
{\sl
there is a unique subspace ${\cal E}_2={\cal E}_1^\bot$ such that, ${\cal E}_1$ and ${\cal E}_1^\bot$ are
complementary and orthogonal with respect to the Euclidean inner product $(\,,\,)$,
$${\cal E}={\cal E}_1\oplus{\cal E}_1^\bot\quad{\rm and}
\quad({\cal E}_1\,,\,{\cal E}_1^\bot)=0\quad\Rightarrow
\quad\forall\,u\in {\cal E}_1,\ \forall\,v\in {\cal E}_1^\bot\quad(u\,,\,v)=0\,. \eqno(2.2)$$
Then, ${\cal E}_1^\bot$, is called the {\it orthogonal (with respect to $(\,,\,))$, complement} of ${\cal E}_1$
}
\note{The uniqueness of ${\cal E}_1^\bot$ follows from the non-degenerate character of the inner product.
See, e.g., [9] section 62 and the {\it Projection theorem} in section 66, [10]  section 1 proposition 2 or [11] 3.3-3 and 3.3-4.
In [8] chapter VII, section 4, complementary and orthogonal subspaces are said to be `complementary'.}.

A constructive method to define ${\cal E}_1^\bot$ is to make use of the {\it Gram-Schmidt orthogonalization}
(G-SO)
\note{See, e.g., [9] section 65, [10] chapter XIV, section 7, [11] 3.4-6 or [14] 10.1 (7).}.
This is a canonical inductive procedure of getting an orthogonal basis $\{E_1,\ldots,E_N\}$ of ${\cal E}$,
starting with an arbitrary basis $\{e_1,\ldots,e_N\}$.
For $b_{j,j},\,j=1,\ldots,n$ some arbitrary real nonzero finite factors, one defines
recurrently
\note{If the inner product were not definite, an arbitrary basis vector as $e_1$ might be
such that $(e_1\,,\,e_1)$ vanishes.
But the non-degenerate character of the inner product implies that any basis vector, say $E_1$, of an
orthogonal basis has to be such that $(E_1\,,\,E_1)\ne0$, otherwise this nonzero vector would be
orthogonal to all vectors in contradiction with the hypothesis.},
$$\eqalignno{
E_1 &:=e_1 b_{1,1}^{-1} \cr
E_2 &:=(-E_1 b_{1,2}+e_2) b_{2,2}^{-1} \cr
\hbox to 0.5cm{\dotfill} &\phantom{=}\hbox to 4cm{\dotfill} \cr
E_n &:=(-E_1 b_{1,n}-\cdots-E_{n-1} b_{n-1,n}+e_n) b_{n,n}^{-1}\quad n=1,\ldots,N &(2.3)\cr
}$$
such that,
$$(E_j\,,\,E_k)=h_j \delta_{j,k}\quad j,k=1,\ldots,n
\qquad E_n=\sum_{m=1}^n e_m a_{m,n}\quad e_n=\sum_{m=1}^n E_m b_{m,n}\quad a_{n,n}=b_{n,n}^{-1}\,. \eqno(2.4)$$
Once a basis is chosen, the inner product is characterized by the {\it metric tensor} components,
$$g_{j,k}:=(e_j\,,\,e_k)\quad i,j=1,\ldots,N\,. \eqno(2.5)$$
It is known
\note{This is usually considered within the framework of orthogonal functions and especially polynomials.
See, e.g., [12] (2.2.6) or
[14] 10.1 (9), (10), 10.3 (3) and (4), with $k_n:=b_{n,n}^{-1}$, $G_n:=Z_n$ and $c_{j+k}:=g_{j,k}$.}
that $E_n$ can be written formally in terms of a determinant as,
$$E_n={b_{n,n}^{-1}\over Z_{n-1}} \det\left[\matrix{
\bigl(g_{j,k}\bigr) \cr
\noalign{\smallskip}
\bigl(e_k\bigr) \cr
}\right]_{j=1,\ldots,n-1\atop k=1,\ldots,n\hfill}\quad  n=1,\ldots,N \eqno(2.6)$$
where $Z_n$ is the nonzero Gram determinant
\note{The inner product being positive definite, the real symmetric matrix $(g_{j,k})_{j,k=1,\ldots,N}$
is positive definite.
All its {\it principal minors} are positive, see, e.g. [14] section 6.2.4.
Hence $Z_n\ne0$ for $n=0,\ldots,N$.}
$$Z_0:=1\qquad Z_n:=\det\bigl[g_{j,k}\bigr]_{j,k=1,\ldots,n}\quad n=1,\ldots,N\,. \eqno(2.7)$$
Actually, {\sl all the relevant quantities can be expressed in terms of determinants with $g_{j,k}$ as entries,
except possibly for the last row}.
One has for $n=1,\ldots,N$,
$$\eqalignno{
h_n &=b_{n,n}^{-1} (e_n\,,\,E_n)
={b_{n,n}^{-2}\over Z_{n-1}} \det\left[\matrix{
\bigl(g_{j,k}\bigr) \cr
\noalign{\smallskip}
\bigl(g_{n,k}\bigr) \cr
}\right]_{j=1,\ldots,n-1\atop k=1,\ldots,n\hfill}
=b_{n,n}^{-2} {Z_n\over Z_{n-1}} &(2.8)\cr
a_{m,n} &\phantom{:}={b_{n,n}^{-1}\over Z_{n-1}} \det\left[\matrix{
\bigl(g_{j,k}\bigr) \cr
\noalign{\smallskip}
\bigl(\delta_{m,k}\bigr) \cr
}\right]_{j=1,\ldots,n-1\atop k=1,\ldots,n\hfill}
=(-1)^{m+n} {b_{n,n}^{-1}\over Z_{n-1}} \det\bigl[g_{j,k}\bigr]_{j=1,\ldots,n-1
\atop{k=1,\ldots,n\atop k\ne m\hfill}\hfill}\,. &(2.9)
}$$
It follows from equation (2.3) that for $m=1,\ldots,n-1$, the orthogonality relation $(E_m\,,\,E_n)=0$ yields
$h_m b_{m,n}=(E_m\,,\,e_n)$.
Then, using equations (2.6) and (2.8) and the symmetry property, $g_{k,n}=g_{n,k}$,
the connection coefficient $b_{m,n}$ reads for $1\le m\le n<N$,
$$b_{m,n}={b_{m,m}^{-1}\over h_m Z_{m-1}} \Bigl(\det\left[\matrix{
\bigl(g_{j,k}\bigr) \cr
\noalign{\smallskip}
\bigl(e_k\bigr) \cr
}\right]_{j=1,\ldots,m-1\atop k=1,\ldots,m\hfill}\,,\,e_n\Bigr)
={b_{m,m}\over Z_m} \det\left[\matrix{
\bigl(g_{j,k}\bigr) \cr
\noalign{\smallskip}
\bigl(g_{n,k}\bigr) \cr
}\right]_{j=1,\ldots,m-1\atop k=1,\ldots,m\hfill}\,. \eqno(2.10)$$
This formula is not given in the standard textbooks already quoted.

There is an infinity of orthogonal bases of the subspace generated by the $n$ linearly
independent vectors $e_1,\ldots,e_n$
\note{The sequence $\{E'_1,\ldots,E'_n\}$ is also an orthogonal basis with the same normalization (2.4)
if and only if it is obtained from $\{E_1,\ldots,E_n\}$ by an isometric linear mapping,
$E'_j:=\sum_{k=1}^n E_k h_k^{-1/2} {\cal O}_{k,j} h_j^{1/2},\,j=1,\ldots,n$
where, $({\cal O}_{k,j})_{k,j=1,\ldots,n}$ is any $n\times n$ orthogonal matrix in the orthogonal group
$O(n)$.}.
The G-SO process shows that:
{\sl
for any set of nonzero finite factors
$\{b_{j,j}\ne0,\,j=1,\ldots,n\}$, there is a unique sequence of orthogonal basis vectors $\{E_1,\ldots,E_n\}$
satisfying equation (2.4), i.e. such that for $m=1,\ldots,n$, $E_m$ is a linear combination of $e_1,\ldots,e_m$
with the nonzero component $b_{m,m}^{-1}$ on $e_m$.
}

Now, assuming $\{e_1,\ldots,e_{N_1}\}$ is a basis of the subspace ${\cal E}_1$ and applying G-SO up to $n=N_1$,
then $\{E_1,\ldots,E_{N_1}\}$ is an orthogonal basis of ${\cal E}_1$.
Carrying on this procedure up to $n=N$, the $N_2$ orthogonal basis vectors
$E_{N_1+1},\ldots,E_{N}$ determine the unique subspace ${\cal E}_1^\bot$ satisfying equation (2.2).

In the commonly used {\it Dirac notation}
\note{See, e.g., [8] chapter VII.},
each vector $v$ in ${\cal E}$ is denoted as the {\it ket} $|v\rangle$.
Each element of the {\it dual space} ${\cal E}^*$ is denoted as a {\it bra} $\langle u|$.
Then, the {\it orthogonality} and the {\it closure relations} for the basis $\{E_1,\ldots,E_N\}$ read
$$\langle E_j|E_k\rangle=h_j \delta_{j,k}\quad j,k=1,\ldots,N
\qquad\sum_{n=1}^N |E_n\rangle\,h_n^{-1}\,\langle E_n|=I_{\cal E} \eqno(2.11)$$
where $I_{\cal E}$ denotes the identity operator on ${\cal E}$.
The projector $\Pi_{{\cal E}_1}$ onto ${\cal E}_1$ along ${\cal E}_1^\bot$, and the
projector $\Pi_{{\cal E}_1^\bot}$ onto ${\cal E}_1^\bot$ along ${\cal E}_1$
are said to be {\it orthogonal projectors} or {\it perpendicular projectors with respect to
$(\,,\,)$}.
One has
$$\Pi_{{\cal E}_1}=\sum_{n=1}^{N_1} |E_n\rangle\,h_n^{-1}\,\langle E_n|
\qquad\Pi_{{\cal E}_1^\bot}=\sum_{n=N_1+1}^N |E_n\rangle\,h_n^{-1}\,\langle E_n|
\qquad\Pi_{{\cal E}_1}+\Pi_{{\cal E}_1^\bot}=I_{\cal E}\,. \eqno(2.12)$$
Although often convenient, {\it this compact notation must be handled with care in the continuation,
since several inner products are going to be considered simultaneously}, see equation (2.20).
Therefore, the Dirac notation will be used henceforth only when dealing with orthogonal projectors.

\medskip
\noindent {\bf Remark.}
Following the idea of `constraint' considered by B Giraud {\it et al} for equation (1.1), or more
generally (1.17), ${\cal E}_2={\cal E}_1^\bot$ can be alternately defined as the subset of vectors $v$ in
${\cal E}$ satisfying the $i=N_1$ constraints, $(e_j\,,\,v)=0,\,j=1,\ldots,N_1$, where $e_j,\,j=1,\ldots,N_1$
are linearly independent vectors (a basis of ${\cal E}_1$ in our first formulation).
Since these constraints are linear in $v$, ${\cal E}_2$ is a subspace.
More precisely here, they are associated with the linear forms,
$\langle e_j|,\,j=1,\ldots,N_1$, and read $\langle e_j|v\rangle=0$.
It follows that $v$ belongs to the kernels of each of these forms, each kernel being of codimension one
\note{Due to the non-degenerate character of the inner product, the kernel of $\langle e_j|$ is the
$(N-1)$-dimensional hyperplane orthogonal to $e_j$ with respect to $(\,,\,)$ in the $N$-dimensional space
${\cal E}$.}.
Now, the non-degenerate character of the inner product implies that these $N_1$ forms are
linearly independent.
Consequently, ${\cal E}_2$ is a subspace of codimension $N_1$.

In a more general problem, independently of the existence of an inner product, ${\cal E}_2$ could be defined
as the intersection of the kernels of $N_1$ linearly independent linear forms $f_j,\,j=1,\ldots,N_1$
in ${\cal E}^*$.
For any given basis $\{e_j,\,j=1,\ldots,N\}$ of ${\cal E}$, these forms are characterized by the coefficients
$f_{j,k}:=f_j(e_k),\,j,k=1,\ldots,N$.
Then, with $v=\sum_{j=1}^N v_j e_j$, the subspace ${\cal E}_2$ is determined by the $N_1$ linear equations
with $N$ unknowns, $\sum_{k=1}^N f_{j,k} v_k=0,\,j=1,\ldots,N_1$.
The rank of this system being $N_1$, the solution $v$, defining ${\cal E}_2$ uniquely, depends linearly on $N-N_1$
parameters, e.g. $v_{N_1+1},\ldots,v_N$ assuming $\det[f_{j,k}]_{j,k=1,\ldots,N_1}\ne0$
\note{When the linear forms are defined by $f_j:=\langle e_j|,\,j=1,\ldots,N_1$, as in the case we consider,
then $f_{j,k}=g_{j,k}:=(e_j\,,\,e_k)$ and thus $Z_{N_1}\ne0$, see footnote 21.}.

Nevertheless, the physical applications quoted in section 1 involve constraints which
are particular linear forms, i.e. bras associated with an Euclidean inner product $(\,,\,)$.
Only this case is considered in this paper, studying below the interplay of the initial inner product $(\,,\,)$
with new inner products, $(\,,\,)_1$ on ${\cal E}_1$, and mainly $(\,,\,)_2$ on ${\cal E}_2$.

\bigskip
\noindent{\it 2.2. Two block orthogonal subspaces and bases}

\medskip
\noindent {\it 2.2.1. Definitions and properties.}
From above, for any subspace ${\cal E}_1$ of an Euclidean space $\cal E$, with the inner
product $(\,,\,)$, the complementary and orthogonal subspace ${\cal E}_1^\bot$ is uniquely determined.
Now, each of these subspaces can be endowed with a new Euclidean structure by any new Euclidean
inner product $(\,,\,)_1$ on ${\cal E}_1$ and $(\,,\,)_2$ on ${\cal E}_1^\bot$
(these inner products can be possibly defined on the whole space ${\cal E}$).
Then, ${\cal E}_1$ and ${\cal E}_1^\bot$ are said to be {\it block orthogonal} (BO) {\it subspaces}.

Every vector in $\cal E$ having a unique decomposition on the complementary subspaces ${\cal E}_1$ and
${\cal E}_1^\bot$, these inner products induce a new Euclidean structure on $\cal E$ with the
inner product $(\,,\,)_0$ defined by
$${\forall u=u_1+u_2\in{\cal E}\quad u_1\in{\cal E}_1,\ u_2\in{\cal E}_1^\bot
\atop\forall v=v_1+v_2\in{\cal E}\quad v_1\in{\cal E}_1,\ v_2\in{\cal E}_1^\bot}
\qquad(u\,,\,v)_0:=(u_1\,,\,v_1)_1+(u_2\,,\,v_2)_2 \eqno(2.13)$$
or, equivalently, ${\cal E}_1$ and ${\cal E}_1^\bot$ being orthogonal with respect to $(\,,\,)$,
$$(u\,,\,v)_0:=\left\{\matrix{
(u\,,\,v)_1\hfill &\forall\,u,v\in{\cal E}_1\hfill\cr
\noalign{\smallskip}
(u\,,\,v)=0\hfill &\forall\,u\in{\cal E}_1\ \forall\,v\in{\cal E}_1^\bot\cr
\noalign{\smallskip}
(u\,,\,v)_2\hfill &\forall\,u,v\in{\cal E}_1^\bot\,.\hfill\cr
}\right. \eqno(2.14)$$
Now, an orthogonal basis $\{\phi_1,\ldots,\phi_N\}$ of $\cal E$ with respect to $(\,,\,)_0$, i.e. such that,
$$(\phi_j\,,\,\phi_k)_0=H_j \delta_{j,k}\quad j,k=1,\ldots,N \eqno(2.15)$$
or, from equation (2.14),
$$\left\{\matrix{
\hfill(\phi_j\,,\,\phi_k)_1\ &=H_j \delta_{j,k}\hfill &j,k=1,\ldots,N_1\hfill\cr
\noalign{\smallskip}
\hfill(\phi_j\,,\,\phi_{N_1+k}) &=0\hfill &j=1,\ldots,N_1\quad k=1,\ldots,N-N_1\hfill\cr
\noalign{\smallskip}
\hfill(\phi_{N_1+j}\,,\,\phi_{N_1+k})_2 &=H_{N_1+j} \delta_{j,k}\hfill &j,k=1,\ldots,N-N_1\hfill\cr
}\right. \eqno(2.16)$$
can be obtained using G-SO again, separately in ${\cal E}_1$ and ${\cal E}_1^\bot$.
Then, $\{\phi_1,\ldots,\phi_{N_1}\}$ and $\{\phi_{N_1+1},\ldots,\phi_N\}$ are said to be
{\it BO bases}.

It should be noted that

\noindent
 - the inner product $(\,,\,)_0$ defined on ${\cal E}$ by equation (2.13) depends on
${\cal E}_1$ and on the three inner products $(\,,\,)$, $(\,,\,)_1$ and $(\,,\,)_2$;

\noindent - an orthogonal basis of ${\cal E}_1$ with respect to $(\,,\,)_1$, $\{\phi_1,\ldots,\phi_{N_1}\}$,
depends only on this inner product.
It can be determined in the standard way, using G-SO to orthogonalize any basis
$\{e_1,\ldots,e_{N_1}\}$ of ${\cal E}_1$ (possibly an orthogonal basis with respect to $(\,,\,)$).
Thereby, with arbitrary real nonzero finite factors $b_{n,n},\,n=1,\ldots,N_1$, one has
$$\eqalignno{
&\{e_1,\ldots,e_{N_1}\}\ \buildrel\hbox{G-SO}\over{\hbox to 15mm{\rightarrowfill}}\ \{\phi_1,\ldots,\phi_{N_1}\}\cr
&g_{1;j,k}:=(e_j\,,\,e_k)_1\qquad(\phi_j\,,\,\phi_k)_1=H_j \delta_{j,k}\quad j,k=1,\ldots,N_1 &(2.17)\cr
}$$
applying equations (2.4)--(2.10) with $\phi_n$, $H_n$ and $g_{1;j,k}$ instead of $E_n$, $h_n$ and $g_{j,k}$,
respectively;

\noindent - on the other hand, an orthogonal basis of ${\cal E}_1^\bot$ with respect to $(\,,\,)_2$,
$\{\phi_{N_1+1},\ldots,\phi_N\}$, depends on ${\cal E}_1$ (regardless to the inner product $(\,,\,)_1$),
and on both inner products $(\,,\,)$ and $(\,,\,)_2$.
It can be obtained applying G-SO twice:

\noindent (i) once to determine ${\cal E}_1^\bot$ itself by the orthogonal basis with respect to $(\,,\,)$,
$\{E_{N_1+1},\ldots,E_N\}$.
With arbitrary real nonzero finite factors $b_{N_1+n,N_1+n},\,n=1,\ldots,N-N_1$, one has readily from equations
(2.4)--(2.10),
$$\eqalignno{
&\{e_1,\ldots,e_N\}\ \buildrel\hbox{G-SO}\over{\hbox to 15mm{\rightarrowfill}}
\ \{E_1,\ldots,E_{N_1},E_{N_1+1},\ldots,E_N\}\cr
&g_{j,k}:=(e_j\,,\,e_k)\qquad(E_j\,,\,E_k)=h_j \delta_{j,k}\quad j,k=1,\ldots,N\,; &(2.18)\cr
}$$

\noindent (ii) once again to orthogonalize $\{E_{N_1+1},\ldots,E_N\}$ with respect to $(\,,\,)_2$.
Thereby, with arbitrary real nonzero finite factors $\beta_{N_1+n,N_1+n},\,n=1,\ldots,N-N_1$, one has
$$\eqalignno{
&\{E_{N_1+1},\ldots,E_N\}\ \buildrel\hbox{G-SO}\over{\hbox to 15mm{\rightarrowfill}}
\ \{\phi_{N_1+1},\ldots,\phi_N\}\cr
&g_{2;j,k}:=(E_{N_1+j}\,,\,E_{N_1+k})_2\qquad(\phi_{N_1+j}\,,\,\phi_{N_1+k})_2=H_{N_1+j} \delta_{j,k}
\quad j,k=1,\ldots,N-N_1 &(2.19)\cr
}$$
applying equations (2.4)--(2.10)  with $\phi_{N_1+n}$, $H_{N_1+n}$ and $g_{2;j,k}$
instead of $E_n$, $h_n$ and $g_{j,k}$, respectively.

\smallskip
{\sl There is an infinity of BO bases.}
They are related by isometric linear mappings within each one of the subspaces ${\cal E}_1$ and ${\cal E}_1^\bot$,
see footnote 22.

\medskip
\noindent {\it 2.2.2. Projection operators.}
To use the Dirac notation, {\it one has to take care of the dependence of the one-to-one
mapping between the vector space and its dual space upon the inner product considered}.
Thus, to each ket $|u\rangle$, there corresponds now four distinct bras, denoted $\langle u|$, $_0\langle u|$,
$_1\langle u|$ and $_2\langle u|$, and defined by,
$$\forall\,|v\rangle\in{\cal E}\quad\langle u|v\rangle:=(u\,,\,v)
\qquad_j\langle u|v\rangle:=(u\,,\,v)_j\quad j=0,1,2\,. \eqno(2.20)$$
The projectors orthogonal with respect to $(\,,\,)$, $\Pi_{{\cal E}_1}$ onto ${\cal E}_1$ and
$\Pi_{{\cal E}_1^\bot}$ onto ${\cal E}_1^\bot$, can be readily expressed only in terms of bases of ${\cal E}$
which can be split into a basis of ${\cal E}_1$ and a basis of ${\cal E}_1^\bot$.
This is the case for the basis $\{E_1,\ldots,E_N\}$ orthogonal with respect to $(\,,\,)$.
Then, the projectors are given by equation (2.12) in terms of the bras $\langle\ |$.
This is also the case with the BO basis  $\{\phi_1,\ldots,\phi_N\}$.
Then, with the bras $_0\langle\ |$ associated with $(\,,\,)_0$, the orthogonality relation (2.15)
and the closure relation read
$$_0\langle\phi_j|\phi_k\rangle=H_j \delta_{j,k}\quad j,k=1,\ldots,N
\qquad\sum_{j=1}^N |\phi_j\rangle\,H_j^{-1}\ _0\langle\phi_j|=I_{\cal E}\,. \eqno(2.21)$$
The projectors $\Pi_{{\cal E}_1}$ and $\Pi_{{\cal E}_1^\bot}$ are given by
$$\Pi_{{\cal E}_1}=\sum_{j=1}^{N_1} |\phi_j\rangle\,H_j^{-1}\ _0\langle\phi_j|
\qquad\Pi_{{\cal E}_1^\bot}=\sum_{j=N_1+1}^N |\phi_j\rangle\,H_j^{-1}\ _0\langle\phi_j|\,. \eqno(2.22)$$
Now, one may ask to express these projectors in terms of the inner product $(\,,\,)_2$ and one of its
orthogonal basis.
Assuming the inner product $(\,,\,)_2$ is defined on the whole space ${\cal E}$,
let us choose the basis $\{E_{2;1},\ldots,E_{2;N}\}$ one gets
applying the G-SO procedure on the basis $\{E_1,\ldots,E_N\}$, i.e. such that,
$$\eqalignno{
\{E_1,&\ldots,E_N\}\ \buildrel\hbox{G-SO}\over{\hbox to 15mm{\rightarrowfill}} {\ \{E_{2;1},\ldots,E_{2;N}\}}
\qquad(E_{2;j}\,,\,E_{2;k})_2=h_{2;j} \delta_{j,k}\quad j,k=1,\ldots,N\cr
E_{2;n} &=\sum_{m=1}^n E_m a_{2;m,n}\qquad E_n=\sum_{m=1}^n E_{2;m} b_{2;m,n}
\quad n=1,\ldots,N &(2.23)\cr
}$$
where, $a_{2;n,n}=b_{2;n,n}^{-1},\,n=1,\ldots,N$ are some arbitrary real nonzero finite factors.
Then, the orthogonality and the closure relations in ${\cal E}$ read
$$_2\langle E_{2;j}|E_{2;k}\rangle=h_{2;j} \delta_{j,k}\quad j,k=1,\ldots,N
\qquad\sum_{n=1}^N |E_{2;n}\rangle\,h_{2;n}^{-1}\ _2\langle E_{2;n}|=I_{\cal E}\,. \eqno(2.24)$$
From the closure relations (2.11) and (2.24), one has
$$\eqalignno{
|E_{2;n}\rangle &=\sum_{m=0}^n |E_m\rangle\,h_m^{-1}\,\langle E_m|E_{2;n}\rangle
\qquad\qquad h_m^{-1}\,\langle E_m|E_{2;n}\rangle=a_{2;m,n} &(2.25)\cr
|E_n\rangle &=\sum_{m=0}^n |E_{2;m}\rangle\,h_{2;m}^{-1}\ _2\langle E_{2;m}|E_n\rangle
\qquad h_{2;m}^{-1}\ _2\langle E_{2;m}|E_n\rangle=b_{2;m,n} &(2.26)\cr
}$$
where, using equation (2.23), the upper bound of the sums over $m$ are set equal to $n$ instead of $N$.
In other words, with the choice of basis $\{E_{2;1},\ldots,E_{2;N}\}$, $\langle E_m|E_{2;n}\rangle$ and
$_2\langle E_{2;m}|E_n\rangle$ vanish for $m>n$.
Finally, using equations (2.12) and (2.24)--(2.26), the projectors $\Pi_{{\cal E}_1}$ and $\Pi_{{\cal E}_1^\bot}$
can be expressed in terms of the bras $_2\langle\ |$ as
$$\eqalignno{
\Pi_{{\cal E}_1} &=\sum_{j=1}^n \sum_{k=n}^N |E_{2;j}\rangle
\,\Bigl(\sum_{n=1}^{N_1} b_{2;j,n} a_{2;n,k}\Bigr) h_{2;k}^{-1}
\ _2\langle E_{2;k}| &(2.27)\cr
\noalign{\smallskip}
\Pi_{{\cal E}_1^\bot}
&=\sum_{n=1}^N h_{2;n}^{-1}\,|E_{2;n}\rangle\ _2\langle E_{2;n}|-\Pi_{{\cal E}_1}\,. &(2.28)\cr
}$$

\bigskip
\noindent{\it 2.3. Extension to more than two subspaces}

\medskip
\noindent {\it 2.3.1. First generalization.}
One may ask wether it is possible to consider three, or more, BO subspaces.
From above, for any given $N_1$-dimensional subspace ${\cal E}_1$ of the Euclidean $N_{1,2}$-dimensional space
${\cal E}_{1,2}$ with the inner product $(\,,\,)_{1,2}$, there is a unique complementary and orthogonal
$(N_{1,2}-N_1)$-dimensional subspace ${\cal E}_2$ such that, according to equation (2.2),
${\cal E}_{1,2}={\cal E}_1\oplus{\cal E}_2$ and $({\cal E}_1\,,\,{\cal E}_2)_{1,2}=0$.
Now, let us assume ${\cal E}_{1,2}$ is a subspace of an Euclidean $N$-dimensional space $\cal E$
with the inner product $(\,,\,)$.
Iterating the procedure above, there is a unique complementary and orthogonal $(N-N_{1,2})$-dimensional subspace
${\cal E}_3$, such that ${\cal E}={\cal E}_{1,2}\oplus{\cal E}_3$ and
$({\cal E}_{1,2}\,,\,{\cal E}_3)=0$.
Thereby, both ${\cal E}_1$ and ${\cal E}_2$ are orthogonal to ${\cal E}_3$ with respect to the {\it same} inner
product $(\,,\,)$.

More generally, for any given subspace ${\cal E}_1\oplus{\cal E}_2$ of $\cal E$
(e.g., such that $({\cal E}_1\,,\,{\cal E}_2)_{1,2}=0$) and two
{\it distinct} Euclidean inner products $(\,,\,)_{1,3}$ and $(\,,\,)_{2,3}$ defined on $\cal E$,
does there exist a subset ${\cal E}_3$ such that,
$${\cal E}={\cal E}_1\oplus{\cal E}_2\oplus{\cal E}_3\quad{\rm and}
\quad\left\{\matrix{
({\cal E}_1\,,\,{\cal E}_3)_{1,3}=0\phantom{\,.} \cr
\noalign{\smallskip}
({\cal E}_2\,,\,{\cal E}_3)_{2,3}=0\,? \cr
}\right. \eqno(2.29)$$
From the linearity of the inner product, if it exists, ${\cal E}_3$ is a subspace.
Let us try to determine ${\cal E}_3$ by constructing one of its basis.
Let $\{e_1,\ldots,e_N\}$ be any basis of $\cal E$ such that $\{e_1,\ldots,e_{N_1}\}$ and
$\{e_{N_1+1},\ldots,e_{N_1+N_2}\}$ are bases of ${\cal E}_1$ and ${\cal E}_2$, respectively.
Now, setting
$$\varepsilon_n:=\sum_{m=1}^{N_1+N_2}e_m c_{m,n}+e_{N_1+N_2+n}\quad n=1,\ldots,N-(N_1+N_2) \eqno(2.30)$$
these vectors are linearly independent.
For $n$ given, the $N_1+N_2$ coefficients $c_{m,n},\,m=1,\ldots,N_1+N_2$ have to satisfy the $N_1+N_2$
orthogonality conditions of $\varepsilon_n$ with ${\cal E}_1$ and ${\cal E}_2$ according to equation (2.29),
$$\left\{\matrix{
\hfill(e_j\,,\,\varepsilon_n)_{1,3}=0\quad j=1,\ldots,N_1\phantom{\,.} \cr
\noalign{\smallskip}
(e_{N_1+j}\,,\,\varepsilon_n)_{2,3}=0\quad j=1,\ldots,N_2\,. \cr
}\right. \eqno(2.31)$$
With the metric tensor components, $g_{\sigma,3;j,k}:=(e_j\,,\,e_k)_{\sigma,3},\,j,k=1,\ldots,N$
for $\sigma=1,2$ and,
$${\bf G}:=\pmatrix{
\bigl(g_{1,3;j,k}\bigr)_{j=1,\ldots,N_1\hfill\atop k=1,\ldots,N_1+N_2\hfill}\hfill \cr
\noalign{\smallskip}
\bigl(g_{2,3;N_1+j,k}\bigr)_{j=1,\ldots,N_2\hfill\atop k=1,\ldots,N_1+N_2\hfill}\hfill \cr
}
\qquad{\bf V_n}:=\pmatrix{
\bigl(g_{1,3;j,N_1+N_2+n}\bigr)_{j=1,\ldots,N_1}\hfill \cr
\noalign{\smallskip}
\bigl(g_{2,3;N_1+j,N_1+N_2+n}\bigr)_{j=1,\ldots,N_2}\hfill \cr
} \eqno(2.32)$$
the linear equations (2.31) read in matrix notation,
${\bf G}\,\bigl(c_{m,n}\bigr)_{m=1,\ldots,N_1+N_2}=-{\bf V_n}$.
Now, $\det{\bf G}$ is not a Gram determinant, as it would be if $(\,,\,)_{1,3}=(\,,\,)_{2,3}$.
Therefore, $\bf G$ may be a singular matrix, and the linear equations (2.31) have a solution if
and only if the rank of $\bf G$ and the rank of the so-called {\it augmented matrix} $({\bf G V_n})$,
obtained by adjoining to $\bf G$ the column ${\bf V_n}$, are the same, say $r$.
Then, the solutions for $\varepsilon_n$ depend linearly on $N_1+N_2-r$ parameters.
Since the principal minors $\det[g_{1,3;j,k}]_{j,k=1,\ldots,N_1}$ and
$\det[g_{2,3;N_1+j,N_1+k}]_{j,k=1,\ldots,N_2}$ are Gram determinants, one has
$\sup(N_1,N_2)\le r\le N_1+N_2$.
Thereby, {\sl the problem of the existence of ${\cal E}_3$ satisfying equation (2.29), may have a unique solution,
no solution or an infinite number of solutions}.
That these three cases do happen has still to be proven.
In other words, it has to be checked that the relations required between the two metric tensors are compatible
with their symmetric and positive-definite character.
Examples with polynomials and $N_1=N_2=1$ and $N=3$ are given in appendix B.
In the generic case, the rank of $\bf G$ is $N_1+N_2$, i.e. $\det{\bf G}\ne0$,
and  ${\cal E}_3$ is defined uniquely by the basis vectors $\varepsilon_1,\ldots,\varepsilon_{N-N_1-N_2}$.
Thus, in this case, for given ${\cal E}_1\subset{\cal E}_{1,2}\subset{\cal E}$ and the Euclidean inner products
$(\,,\,)_{1,2}$, $(\,,\,)_{1,3}$ and $(\,,\,)_{2,3}$, there are {\it unique} subspaces ${\cal E}_2$ and
${\cal E}_3$ satisfying equation (2.29) with ${\cal E}_{1,2}={\cal E}_1\oplus{\cal E}_2$.
Now, endowing  ${\cal E}_1$, ${\cal E}_2$ and ${\cal E}_3$ with new Euclidean inner products $(\,,\,)_{1,1}$,
$(\,,\,)_{2,2}$ and $(\,,\,)_{3,3}$, respectively, these three subspaces are said to be BO.
A new Euclidean structure on $\cal E$ is induced with the inner product
defined as in equation (2.14) by,
$(u\,,\,v)_0:=(u\,,\,v)_{\sigma,\sigma}$ if $u,v\in{\cal E}_\sigma,\,\sigma=1,2,3$, and $0$ otherwise.
Orthogonal bases of ${\cal E}_\sigma$ with respect to $(\,,\,)_{\sigma,\sigma}$ for $\sigma=1,2,3$, respectively,
are said to be BO bases
\note{This study can be extended, with similar conclusions, to $p>2$ subspaces
${\cal E}_\sigma$ and inner products $(\,,\,)_{\sigma,p+1}$, not all the same, defined on $\cal E$ for
$\sigma=1,\ldots,p$.
Then, the question is to determine the subspace ${\cal E}_{p+1}$  such that
${\cal E}=\oplus_{\sigma=1}^{p+1}\,{\cal E}_\sigma$ and
$({\cal E}_\sigma\,,\,{\cal E}_{p+1})_{\sigma,p+1}=0,\ \sigma=1,\ldots,p$.}.

\medskip
\noindent {\it 2.3.2. Second generalization.}
Let ${\cal E}$ be an $N$-dimensional vector space endowed with two distinct Euclidean inner products $(\,,\,)$
and $(\,,\,)'$.
{\it For a given $N_1$-dimensional subspace ${\cal E}_1$, does their exist a subset ${\cal F}$ of vectors
orthogonal to ${\cal E}_1$ with respect to $(\,,\,)$ and $(\,,\,)'$ at the same time?}
It follows from section 2.1 that ${\cal F}$ is the subspace ${\cal E}_1^\bot\bigcap{{\cal E}_1^\bot}'$,
intersection of both orthogonal complements of ${\cal E}_1$, ${\cal E}_1^\bot$ with respect to $(\,,\,)$
and ${{\cal E}_1^\bot}'$ with respect to $(\,,\,)'$.
Let $\{\varepsilon_1',\ldots,\varepsilon_{N-N_1}'\}$ be any basis of ${{\cal E}_1^\bot}'$.
There exist uniquely defined vectors $u_j\in{\cal E}_1$ and $\varepsilon_j\in{\cal E}_1^\bot$ such that
$\varepsilon_j'=u_j+\varepsilon_j,\,j=1,\ldots,N-N_1$.
Then, $\{\varepsilon_1,\ldots,\varepsilon_{N-N_1}\}$ is a basis of ${\cal E}_1^\bot$ and
for any vector $v$ in ${\cal F}$, one has
$$v=\sum_{j=1}^{N-N_1} \alpha_j \varepsilon_j=\sum_{j=1}^{N-N_1} \alpha_j' \varepsilon_j'
\ \Rightarrow\ \sum_{j=1}^{N-N_1} (\alpha_j-\alpha_j') \varepsilon_j=\sum_{j=1}^{N-N_1} \alpha_j' u_j=0\,.
\eqno(2.33)$$
The last equation above vanishes since the left-hand side is a vector in ${\cal E}_1^\bot$
while the right-hand side is in ${\cal E}_1$.
This implies that the dimension of ${\cal F}$ is $N-N_1-r$, where $r$ is the rank of $\{u_1,\ldots,u_{N-N_1}\}$.
(This rank is intrinsic, i.e. it only depends on ${\cal E}_1^\bot$ and ${{\cal E}_1^\bot}'$, independently of the
choices of bases for these subspaces.)
These $N-N_1$ $u$-vectors belonging to the $N_1$-dimensional subspace ${\cal E}_1$, one has $r\le\inf(N_1,N-N_1)$.
As a first example, if these $u$-vectors are linearly independent, $r=N-N_1$ (requiring $N-N_1\le N_1$)
and the dimension of ${\cal F}$ vanishes, i.e. ${\cal F}=\{0\}$.
As a second example, if ${\cal E}_1^\bot={{\cal E}_1^\bot}'$,
although the two inner products are distinct, all the $u$-vectors vanish, $r=0$ and ${\cal F}={\cal E}_1^\bot$.

\medskip
\noindent {\bf Remark.}
Most of the properties considered in this section 2 can be extended to
Hermitian vector space over the complex field
and also to infinite-dimensional Hilbert spaces, the typical spaces used in
quantum theory
\note{See, e.g., [8] chapter V, section 2 and chapter VII, section 4, [11] chapter 3 or [19] chapter VI.}.

\medskip
\bigskip
\noindent{\elevenrmb 3. Application to vector spaces of polynomials}

\bigskip
\noindent{\it 3.1. Euclidean vector space of polynomials}

\medskip
\noindent Let $\mu$ be a non-constant and non-decreasing real function on the real domain ${\cal D}$
such that all the moments are finite,
$$c_n:=\int_{{\cal D}}d\mu(x)\,x^n<\infty\quad n=0,1,\ldots\,. \eqno(3.1)$$
If $\mu$ is absolutely continuous $d\mu(x)=w(x) dx$ where, $w$ is a nonzero and non-negative weight function.
The associated inner product of real functions $f$ and $g$ in the class of square integrable functions
$L_\mu^2({\cal D})$ is denoted by
$$(f\,,\,g):=\int_{\cal D}d\mu(x)\,f(x) g(x)\,. \eqno(3.2)$$
This inner product is symmetric,  $(f\,,\,g)=(g\,,\,f)$,
and it follows from the hypotheses on $\mu$ that it is positive definite, i.e. $(f\,,\,f)>0$
except for $f=0$, where it vanishes.
Now, for $N$ a positive integer, let ${\cal P}_N$ be the $N$-dimensional vector space over the real fields of real
polynomials of degree at most $N-1$ (by definition, ${\cal P}_0:=\{0\}$)
\note{Let us recall the basic property: the ${\cal P}_j,\,j=0,1,\ldots$ are nested, i.e.
$0\le j\le k,\ {\cal P}_j\subseteq{\cal P}_k$.}.
Any inner product defined as above endows ${\cal P}_N$ with an Euclidean structure.
Then, all the general results considered in section 2 can be applied.
Furthermore, new properties are available in connection with (i) the additional characteristics of the vectors,
i.e. the {\it degree}, possibly the {\it parity} or the {\it monomial} character of polynomials
and basically, a richer algebraic structure, the multiplication endowing the set of polynomials with
the structures of a {\it ring} and of an {\it algebra} over the real field
\note{See, e.g., [9] section 35 or [10] chapter V for a formal study of polynomials.},
and (ii) the definition of the inner product in terms of an integral.
The theory of orthogonal polynomials has been extensively studied
\note{The standard textbook on this subject is [12].
See also, e.g., [13] and [14] chapter X.}.
The G-SO of the monomial basis $\{x^0,x^1,\ldots,x^{N-1}\}$ of ${\cal P}_N$ provides {\sl the unique standard
orthogonal basis of polynomials $\{Q_0,\ldots,Q_{N-1}\}$ such that $Q_n$ is a
polynomial of degree $n$ with a given arbitrary real nonzero factor $k_n$ as coefficient of $x^n$}
\note{There is an infinity of orthogonal polynomials, see footnote 22.
As an example, G-SO of the monomial basis taken in the reverse order,
i.e. $\{e_1:=x^{N-1},\ldots,e_N:=x^0\}$, generates the unique orthogonal basis of polynomials
$\{R_1:=E_1,\ldots,R_N:=E_N\}$ such that $R_j$ is a polynomial of degree at most $N-1$ and in which the lowest
degree monomial is $x^{N-j}$ with a given arbitrary nonzero $a_{N-j,j}$ as coefficient:
$R_j=a_{1,j} x^{N-1}+a_{2,j} x^{N-2}+\cdots+ a_{N-j,j} x^{N-j}$, thereby defining another kind
of standard orthogonal polynomials.
},
$$(Q_j\,,\,Q_k)=h_j \delta_{j,k}\quad{\rm and}\quad Q_j=k_j x^j+{\rm O}(x^{j-1})\,. \eqno(3.3)$$
To avoid the one unit lag between the index of these basis vectors and the degree,
{\it from now on, the indices indicating the degree will start at $0$ instead of $1$}.
This induces some trivial changes in the general formulae (2.2)--(2.10) which are given in appendix A.

{\it Classical orthogonal polynomials} (i.e., Hermite,  Laguerre and Jacobi
polynomials) correspond to particular measures $\{{\cal D},d\mu\}$.
These polynomials arise frequently and have been studied in great detail
\note{See, e.g., [14] sections 10.6--10.13, [15] chapter 22 or [16] section 8.9.}.
Particular SBO polynomials associated with these polynomials (as those considered in [1, 3]
for the Hermite case, and in [2] for the Laguerre case) will be studied in forthcoming papers.

\bigskip
\noindent{\it 3.2. Standard block orthogonal polynomials for two subspaces}

\medskip
\noindent Let us apply the general considerations of sections 2.1 and 2.2 to the following case:
(i) ${\cal E}$ is the vector space of polynomials ${\cal P}_N$ with three Euclidean inner products $(\,,\,)$,
$(\,,\,)_1$ and $(\,,\,)_2$ defined
as in equation (3.2) by the measures $\{{\cal D},d\mu\}$, $\{{\cal D}_1,d\mu_1\}$ and $\{{\cal D}_2,d\mu_2\}$,
respectively;
(ii) ${\cal E}_1$ is {\it any subspace} of ${\cal P}_N$, e.g., defined by {\it any sequence}
$\{e_1,\ldots,e_{N_1}\}$ of $N_1$ linearly independent polynomials in ${\cal P}_N$.
For ${\cal E}_1={\cal P}_0$ or ${\cal P}_N$, the problem is trivial.
For $0<N_1<N$, the general procedure given in section 2.2 can be applied
to get BO bases of ${\cal E}_1$ and ${\cal E}_2$ satisfying the orthogonality relations (2.16).
The first step is to determine the unique subspace ${\cal E}_2={\cal E}_1^\bot$ complementary and orthogonal to
${\cal E}_1$ with respect to the inner product $(\,,\,)$.
For that purpose, the basis $\{e_1,\ldots,e_{N_1}\}$ of ${\cal E}_1$ is completed to get a basis
$\{e_1,\ldots,e_N\}$ of ${\cal P}_N$.
The G-SO of this basis provides a basis $\{E_1,\ldots,E_N\}$ of orthogonal polynomials
with respect to the inner product $(\,,\,)$.
Then, ${\cal E}_1^\bot$ is defined uniquely by the sub-basis $\{E_{N_1+1},\ldots,E_N\}$.
Finally, G-SO of $\{e_1,\ldots,e_{N_1}\}$ (or  of $\{E_1,\ldots,E_{N_1}\}$) with respect to $(\,,\,)_1$ and
of $\{E_{N_1+1},\ldots,E_N\}$ with respect to $(\,,\,)_2$ provides BO bases of ${\cal E}_1$ and ${\cal E}_1^\bot$,
respectively.
It should be noted that {\it this general procedure applies whatever the degree of the polynomials defining
${\cal E}_1$ is.}

{\it Henceforth, let us focus on the special case where, ${\cal E}_1={\cal P}_i$, i.e. the subspace of polynomials
of degree less than $i$, for given $0<i<N$.}
0ne basis of this subspace is $\{x^0,\ldots,x^{i-1}\}$.
Then, following the procedure above, the unique subspace ${\cal P}_{i;N}^\bot:={\cal E}_1^\bot$ such that,
$$\bigl({\cal P}_i\,,\,{\cal P}_{i;N}^\bot\bigr)=0\quad{\rm and}
\quad{\cal P}_N={\cal P}_i\oplus{\cal P}_{i;N}^\bot \eqno(3.4)$$
is defined uniquely by the sub-basis $\{Q_i,\ldots,Q_{N-1}\}$ of standard orthogonal polynomials
with respect to the inner product $(\,,\,)$.
For given arbitrary real nonzero finite $k_n:=b_{n,n}^{-1},\,n=0,\ldots,N-1$, these polynomials,
satisfying equation (3.3), are defined uniquely by equations (A.1)--(A.10).

The inner products $(\,,\,)_1$ and $(\,,\,)_2$ induce a new Euclidean structure on ${\cal P}_N$
with the inner product $(\,,\,)_0$ defined as in equations (2.13) and (2.14),
$${\forall p=p_{i;1}+p_{i;2}\in{\cal P}_N\quad p_{i;1}\in{\cal P}_i,\ p_{i;2}\in{\cal P}_{i;N}^\bot
\atop\forall q=q_{i;1}+q_{i;2}\in{\cal P}_N\quad q_{i;1}\in{\cal P}_i,\ q_{i;2}\in{\cal P}_{i;N}^\bot}
\qquad(p\,,\,q)_0:=(p_{i;1}\,,\,q_{i;1})_1+(p_{i;2}\,,\,q_{i;2})_2\,. \eqno(3.5)$$
The BO bases $\{P_{i,0},\ldots,P_{i;i-1}\}$ of ${\cal P}_i$ and $\{P_{i,i},\ldots,P_{i;N-1}\}$ of
${\cal P}_{i;N}^\bot$ form an orthogonal basis of ${\cal P}_N$ with respect to the inner product
$(\,,\,)_0$ such that, as in equation (2.16),
$$(P_{i;m}\,,\,P_{i;n})_0=\left\{\matrix{
(P_{i;m}\,,\,P_{i;n})_1\hfill &=H_{i;m} \delta_{m,n}\hfill &m,n=0,\ldots,i-1\hfill\cr
\noalign{\smallskip}
(P_{i;m}\,,\,P_{i;n})\hfill &=0\hfill &m=0,\ldots,i-1\quad n=i,\ldots,N-1\hfill\cr
\noalign{\smallskip}
(P_{i;m}\,,\,P_{i;n})_2\hfill &=H_{i;m} \delta_{m,n}\hfill &m,n=i,\ldots,N-1\,.\hfill\cr
}\right. \eqno(3.6)$$
Riquiring that $P_{i;n}$ {\it be a polynomial of exact degree} $n$, the equations above determine
{\it uniquely} these polynomials up to an arbitrary nonzero factor in each polynomial:

\smallskip
\noindent - the $P_{i;n},\,n=0,\ldots,i-1$, which only depend on the inner product $(\,,\,)_1$,
are the standard orthogonal polynomials for the measure $\{{\cal D}_1,d\mu_1\}$ as given in appendix A,
with nothing new.
In what follows we will not be concerned with them;

\smallskip
\noindent - the $P_{i;n},\,n=i,\ldots,N-1$, which depend on $i$ (regardless to the inner product $(\,,\,)_1)$
and on both inner products $(\,,\,)$ and $(\,,\,)_2$, are the {\it standard block orthogonal} (SBO) polynomials.
They can be obtained using G-SO to
orthogonalyse the basis $\{Q_i,\ldots,Q_{N-1}\}$ with respect to the inner product $(\,,\,)_2$.
Indeed, the procedure preserves the degree of the polynomials and one has
$$P_{i;n}=\sum_{m=i}^n Q_m A_{i;m,n}\qquad Q_n=\sum_{m=i}^n P_{i;m} B_{i;m,n}
\qquad A_{i;n,n}=B_{i;n,n}^{-1}\,. \eqno(3.7)$$
Hence, with $\widehat P_{i;n}$ a monic polynomial,
$$P_{i;n}:=K_{i;n}\,\widehat P_{i;n}\qquad
\widehat P_{i;n}:=x^n+\widehat R_{i;n} x^{n-1}+\widehat S_{i;n} x^{n-2}+{\rm O}(x^{n-3}) \eqno(3.8)$$
and equations (3.7) and (A.3) yield
$$\eqalignno{
K_{i;n} &=k_n A_{i;n,n}=(b_{n,n} B_{i;n,n})^{-1} &(3.9)\cr
\noalign{\smallskip}
\widehat R_{i;n} &=\widehat r_n+{k_{n-1}\over k_n} {A_{i;n-1,n}\over A_{i;n,n}} &(3.10)\cr
\widehat S_{i;n} &=\widehat s_n+{k_{n-1}\over k_n} {A_{i;n-1,n}\over A_{i;n,n}}\,\widehat r_{n-1}
+{k_{n-2}\over k_n} {A_{i;n-2,n}\over A_{i;n,n}}\,. &(3.11)
}$$
Then, given arbitrary nonzero finite $B_{i;j,j}$ (or equivalently $K_{i;j}$) for $j=i,\ldots,N-1$, and with
$$\gamma_{j,k}:=(Q_j\,,\,Q_k)_2\quad j,k=i,\ldots,N-1 \eqno(3.12)$$
it follows from equations (2.19) and (2.4)--(2.10), that for $i\le m\le n\le N-1$,
$$\eqalignno{
\widehat P_{i;n} &\phantom{:}=k_n^{-1}\,{1\over Z_{i;n-1}} \det\left[\matrix{
\bigl(\gamma_{j,k}\bigr) \cr
\noalign{\smallskip}
\bigl(Q_k\bigr) \cr
}\right]_{j=i,\ldots,n-1\atop k=i,\ldots,n\hfill}
\qquad \widehat P_{i;i}=k_i^{-1}\,Q_i=\widehat Q_i &(3.13)\cr
\noalign{\smallskip}
Z_{i;n} &:=det\bigl[\gamma_{j,k}\bigr]_{j,k=i,\ldots,n}\qquad\qquad\qquad\quad Z_{i;i-1}:=1 &(3.14)\cr
\noalign{\smallskip}
H_{i;n} &\phantom{:}=K_{i;n}^2\,\widehat H_{i;n}
\qquad \widehat H_{i;n}:=k_n^{-2}\,{Z_{i;n}\over Z_{i;n-1}} &(3.15)\cr
\noalign{\smallskip}
A_{i;m,n} &\phantom{:}=K_{i;n}\,\widehat A_{i;m,n}
\qquad \widehat A_{i;m,n}:=k_n^{-1}\,Z_{i;n-1}\,\det\left[\matrix{
\bigl(\gamma_{j,k}\bigr) \cr
\noalign{\smallskip}
\bigl(\delta_{m,k}\bigr) \cr
}\right]_{j=i,\ldots,n-1\atop k=i,\ldots,n\hfill} &(3.16)\cr
\noalign{\smallskip}
B_{i;m,n} &\phantom{:}=K_{i;m}^{-1}\,\widehat B_{i;m,n}
\qquad\widehat B_{i;m,n}:=k_m\,{1\over Z_{i;m}}\,\det\left[\matrix{
\bigl(\gamma_{j,k}\bigr) \cr
\noalign{\smallskip}
\bigl(\gamma_{n,k}\bigr) \cr
}\right]_{j=i,\ldots,m-1\atop k=i,\ldots,m\hfill}\,. &(3.17)
}$$

For $i=0$, the procedure above amounts to orthogonalize the basis $\{Q_0,\ldots,Q_{N-1}\}$
of ${\cal P}_N$ with respect to $(\,,\,)_2$.
Then, $\widehat P_{0;n}$ is the unique standard monic orthogonal polynomial $\widehat Q_{2;n}$
of exact degree $n$, defined for the measure $\{{\cal D}_2,d\mu_2\}$.
From equation (A.6), one has
$$\eqalignno{
\widehat P_{0;n} &=k_n^{-1}\,{\det\left[\matrix{
\bigl(\gamma_{j,k}\bigr) \cr
\noalign{\smallskip}
\bigl(Q_k\bigr) \cr
}\right]_{j=0,\ldots,n-1\atop k=i,\ldots,n\hfill}\over Z_{0;n-1}}
=\widehat Q_{2;n}:={\det\left[\matrix{
\bigl(c_{2;j+k}\bigr) \cr
\bigl(x^k\bigr) \cr
}\right]_{j=0,\ldots,n-1\atop k=0,\ldots,n}\over\det\bigl[c_{2;j+k}\bigr]_{j,k=0,\ldots,n-1}} &(3.18)\cr
}$$
where
$$c_{2;n}:=\int_{{\cal D}_2}d\mu_2(x)\,x^n\,. \eqno(3.19)$$
Thus, for $i=0$, the general results are still valid with ${\cal P}_0:=\{0\}$ and
${\cal P}_{0;N}^\bot:={\cal P}_N$.
For $i=N$, the problem is trivial with ${\cal P}_{N;N}^\bot=\{0\}$.

Finally, in the $i,n$-plane, the SBO polynomial $\widehat P_{i;n}$ is associated with the
integer coordinate point $i,n$ such that $0\le i\le n$.
These points are located in the sector between the two boundary
half straight lines: the diagonal $i=n\ge0$, along which $\widehat P_{n;n}=\widehat Q_n$;
and the $y$-axis $i=0,\,n\ge0$, along which $\widehat P_{0;n}=\widehat Q_{2;n}$,
where, $\widehat Q_n$ and $\widehat Q_{2;n}$ are the standard monic orthogonal polynomials of exact degree $n$
for the measures $\{{\cal D},d\mu\}$ and $\{{\cal D}_2,d\mu_2\}$, respectively.
At the origin, $\widehat P_{0;0}=1$.

\smallskip
From equations (3.7), (A.3) and (A.4), an expansion of $P_{i;n}$ in terms of the monomials $x^m,\,m=0,\ldots,n$,
as well as the inverse expansion, can be written.
As an example, inverting the finite sums over $m$ and $\ell$, one has
$$P_{i;n}=\sum_{m=i}^n \Bigl(\sum_{\ell=0}^m x^\ell a_{\ell,m}\Bigr) A_{i;m,n}
=\sum_{\ell=0}^n x^\ell C_{i;\ell,n}
\qquad C_{i;\ell,n}:=\sum_{m=\sup(i,\ell)}^n a_{\ell,m} A_{i;m,n}\,. \eqno(3.20)$$
However, we have not been able to get a close form for these connection coefficients $C_{i;\ell,n}$, even
in the special cases associated with the classical polynomials considered in forthcoming papers.

\bigskip
\noindent{\it 3.3. Properties of standard block orthogonal polynomials for two subspaces}

\medskip
\noindent {\it 3.3.1. Projection operators.}
Using the Dirac notation, the general formulae given in sections 2.1 and 2.2.2 apply readily
with only some slight modifications in the notations.
In the space ${\cal P}_N$, the orthogonal bases of the inner products $(\,,\,)$, $(\,,\,)_2$ and
$(\,,\,)_0$ are $\{Q_0,\ldots,Q_{N-1}\}$, $\{Q_{2;0},\ldots,Q_{2;N-1}\}$ and $\{P_{i;0},\ldots,P_{i;N-1}\}$,
respectively.
From equation (3.18), $Q_{2;n}\propto P_{0;n}$, therefore, $\{Q_{2;0},\ldots,Q_{2;N-1}\}$
can be replaced by the basis $\{P_{0;0},\ldots,P_{0;N-1}\}$.
Thus, the orthogonality and the closure relations for these orthogonal bases read for $0\le i<N$,
$$\eqalignno{
&\langle Q_j|Q_k\rangle=h_j \delta_{j,k}
\qquad_2\langle P_{0;j}|P_{0;k}\rangle=H_{0;j} \delta_{j,k}
\qquad_0\langle P_{i;j}|P_{i;k}\rangle=H_{i;j} \delta_{j,k}\quad j,k=0,\ldots,N-1 &(3.21)\cr
&\sum_{j=0}^{N-1} |Q_j\rangle\,h_j^{-1}\,\langle Q_j|
=\sum_{j=0}^{N-1} |P_{0;j}\rangle\,H_{0;j}^{-1}\ _2\langle P_{0;j}|
=\sum_{j=0}^{N-1} |P_{i;j}\rangle\,H_{i;j}^{-1}\ _0\langle P_{i;j}|=I_{{\cal P}_N}\,. &(3.22)\cr
}$$
As chosen in section 2.2.2, the basis $\{P_{0;0},\ldots,P_{0;N-1}\}$ is obtained from
$\{Q_0,\ldots,Q_{N-1}\}$ using the G-SO procedure with respect to $(\,,\,)_2$.
(What is peculiar here, is that all the basis vectors $Q_n$, $P_{0;n}$ and $P_{i;n}$ are polynomials {\it of exact
degree $n$}.)
Therefore, with equation (3.7), one has using the closure relations within ${\cal P}_{n+1}$,
$$\eqalignno{
|P_{0;n}\rangle &=I_{{\cal P}_{n+1}}\,|P_{0;n}\rangle
=\sum_{m=0}^n |Q_m\rangle\,h_m^{-1}\,\langle Q_m|P_{0;n}\rangle
\qquad h_m^{-1}\,\langle Q_m|P_{0;n}\rangle=A_{0;m,n} &(3.23)\cr
|Q_n\rangle &=I_{{\cal P}_{n+1}}\,|Q_n\rangle
=\sum_{m=0}^n |P_{0;m}\rangle\,H_{0;m}^{-1}\ _2\langle P_{0;m}|Q_n\rangle
\qquad H_{0;m}^{-1}\ _2\langle P_{0;m}|Q_n\rangle=B_{0;m,n} &(3.24)\cr
}$$
corresponding to equations (2.25) and (2.26), respectively.
In other words, $\langle Q_m|P_{0;n}\rangle$ and $_2\langle P_{0;m}|Q_n\rangle$ vanish for $m>n$,
or else, the matrices which relate these different bases are triangular.
Following equation (2.27), the projectors, $\Pi_{{\cal P}_i}$ onto ${\cal P}_i$
and $\Pi_{{\cal P}_{i;N}^\bot}$ onto ${\cal P}_{i;N}^\bot$, orthogonal with respect to $(\,,\,)$, are,
$$\eqalignno{
\Pi_{{\cal P}_i} &=\sum_{n=0}^{i-1} |Q_n\rangle\,h_n^{-1}\,\langle Q_n|
=\sum_{n=0}^{i-1} |P_{i;n}\rangle\,H_{i;n}^{-1}\ _0\langle P_{i;n}| &(3.25)\cr
&=\sum_{j=0}^n \sum_{k=n}^{N-1}
|P_{0;j}\rangle\Bigl(\sum_{n=0}^{i-1}B_{0;j,n} A_{0;n,k}\Bigr) H_{0;k}^{-1}\ _2\langle P_{0;k}| &(3.26)\cr
\noalign{\smallskip}
\Pi_{{\cal P}_{i;N}^\bot} &=\sum_{n=i}^{N-1} |Q_n\rangle\,h_n^{-1}\,\langle Q_n|
=\sum_{n=i}^{N-1} |P_{i;n}\rangle\,H_{i;n}^{-1}\ _0\langle P_{i;n}| &(3.27)\cr
&=I_{{\cal P}_N}-\Pi_{{\cal P}_i}=\sum_{n=0}^{N-1} |P_{0;n}\rangle\,H_{0;n}^{-1}\ _2\langle P_{0;n}|
-\Pi_{{\cal P}_i}\,. &(3.28)\cr
}$$
As an example, for $i=1$ one has
\note{This kind of relations are given in [2] (9)--(11) in the special case $i=1$.
Unfortunately, these equations are wrong since the different inner products involved have not been taken
properly into account.
This is especially the case when writing ${\cal P}_N=\sum_{n=1}^N |w_n\rangle\langle w_n|$ which should
correspond to our expression (3.25) in terms of $_0\langle\ |$,
not easy to handle as underlined below.
Furthermore, $\langle r|\sigma_N\rangle$ defined with (9) should be $\exp(-1/r)$ times a zero degree polynomial,
i.e. a constant corresponding to our $|Q_0\rangle$
(in addition, $\langle z_n\rangle$ for the Laguerre polynomials should be $(-1)^n\,2$ instead of $2$).
Thus, $|\sigma_N\rangle$ corresponds to some polynomial $|R_N\rangle$ of degree $N$.
Nevertheless, it happens that in (10) the bra $\langle\sigma_N|$ do corresponds to $\propto\langle Q_0|$
and moreover, the normalization coefficient is such that $\langle Q_0|R_N\rangle=1$, yielding (11).
Yet, although $|R_N\rangle\langle\sigma_N|$ is a projector, {\it it is not an orthogonal projector}.}
$$\eqalignno{
\Pi_{{\cal P}_1} &=|Q_0\rangle\,h_0^{-1}\,\langle Q_0|
=|P_{0;0}\rangle\ B_{0;0,0}\,\sum_{k=0}^{N-1} A_{0;0,k} H_{0;k}^{-1}\ _2\langle P_{0;k}|\,. &(3.29)\cr
}$$

\smallskip
Let us emphasize again that all these equations have to be handled with care since different inner products
are involved.
Moreover, when the action of the bras $\langle\ |$ and $_2\langle\ |$ are expressed with a single integral,
e.g. $\forall\,f,p\,\in\,{\cal P}_N\quad\langle f|p\rangle:=\int_{\cal D}d\mu(x)\,f(x) p(x)$,
{\it this is no longer true for the bra $_0\langle f|$}.
Actually, the only way to compute $_0\langle f|p\rangle$ follows from the definition of $(\,,\,)_0$
by equation (3.5), i.e. it is to split both $f$ and $p$ into their components in ${\cal P}_i$ and
${\cal P}_{i;N}^\bot$. Then, this inner product reads as a sum of two integrals.

\medskip
\noindent {\it 3.3.2. Integral representations.}
The relations between the {\it product of differences}, a {\it Vandermonde determinant}
and a {\it polynomial alternant} of any polynomials $p_n=k_n x^n+{\rm O}(x^{n-1})$
are well-known
\note{See, e.g., [16] 14.311, [18] section 7.1 or [20] equation (C.3).}.
With ${\bf y}_n:=\{y_0,\ldots,y_{n-1}\}$, $\Delta_1({\bf y}_1):=1$ and for $n=2,3\ldots$, one has
$$\Delta_n({\bf y}_n):=\prod_{0\le j<k\le n-1}(y_k-y_j)=det\bigl[(y_j)^k\bigr]_{j,k=0,\ldots,n-1}
={1\over\prod_{j=0}^{n-1} k_j}\,det\bigl[p_k(y_j)\bigr]_{j,k=0,\ldots,n-1}\,.  \eqno(3.30)$$

\smallskip
\noindent (i) {\it Integral representation of} $Z_{i;n}$.
In the integral,
$$I:=\int_{{\cal D}_2}d\mu_2(y_i)\cdots\int_{{\cal D}_2}d\mu_2(y_{n})\,\det\bigl[Q_k(y_j)\bigr]_{j,k=i,\ldots,n}^2
\eqno(3.31)$$
the product of measures is symmetric and each determinant is anti-symmetric in the variables $y_i,\ldots,y_n$.
Therefore, one of the two determinants can be replaced by one of the $(n+1-i)!$ terms of its expansion,
say $\prod_{j=i}^n Q_j(y_j)$, provided one multiplies the result by $(n+1-i)!$.
Thereby, integrating independently over the variables $y_i,\ldots,y_n$ ($y_j$ occurs only in the row $j$),
and using equation (3.12), one gets
$$I=(n+1-i)!\,\det[\gamma_{j,k}]_{j,k=i,\ldots,n}\,. \eqno(3.32)$$
With equation (3.14), this yields the integral representation of $Z_{i;n}$ for $n=i,i+1,\ldots$
\note{This result can also be obtained from the Gram equality.
See, e.g., [18] sections 3.7 and 7.4.},
$$Z_{i;n}={1\over(n+1-i)!} \int_{{\cal D}_2}d\mu_2(y_i)\cdots\int_{{\cal D}_2}d\mu_2(y_{n})
\,\det\bigl[Q_k(y_j)\bigr]_{j,k=i,\ldots,n}^2\,. \eqno(3.33)$$
For $i=0$, replacing $Q_k$ by $y^k$, the same steps as above yield the known integral
representation
\note{See, e.g., [12] (2.2.7) and (2.2.11).},
$$\det\bigl[c_{2;j+k}\bigr]_{j,k=0,\ldots,n}
={1\over(n+1)!} \int_{{\cal D}_2}d\mu_2(y_0)\cdots\int_{{\cal D}_2}d\mu_2(y_n)
\,\Delta_{n+1}({\bf y}_{n+1})^2 \eqno(3.34)$$
where, $c_{2;j}$ is defined by equation (3.19).
Therefore, using equations (3.3) and (3.30), one recovers the relation between the determinants
of the metric tensor matrices for the measure $\{{\cal D}_2,d\mu_2\}$ in the two bases
$\{Q_j,\,j=0,\ldots,n\}$ and $\{x^j,\,j=0,\ldots,n\}$,
$$Z_{0;n}=\det\bigl[\gamma_{j,k}\bigr]_{j,k=0,\ldots,n}
=\Bigl(\prod_{j=0}^n k_j\Bigr)^2 \det\bigl[c_{2;j+k}\bigr]_{j,k=0,\ldots,n}\,.  \eqno(3.35)$$

\smallskip
\noindent (ii) {\it Integral representation of} $\widehat P_{i;n}$.
From equations (3.3) and (3.30), one has
$$\Delta_{n+1}({\bf y}_n,x)=\Delta_n({\bf y}_n) \prod_{j=0}^{n-1}(x-y_j)
={1\over\prod_{j=0}^n k_j} \det\left[\matrix{
\bigl(Q_k(y_j)\bigr) \cr
\noalign{\smallskip}
\bigl(Q_k(x)\bigr) \cr
}\right]_{j=0,\ldots,n-1\atop k=0,\ldots,n}\,. \eqno(3.36)$$
Now, for $j=0,\ldots,n-1$, multiplying the $j$th row of the determinant above by $Q_j(y_j)$
and integrating over $y_j$ the elements of this row with the measure $\{{\cal D},d\mu\}$ for $j=0,\ldots,i-1$
and the measure $\{{\cal D}_2,d\mu_2\}$ for $j=i,\ldots,n-1$, one finds with equations (3.3) and (3.12),
$$\eqalignno{
\int_{\cal D}d\mu(y_0)\cdots &\int_{\cal D}d\mu(y_{i-1})
\int_{{\cal D}_2}d\mu_2(y_i)\cdots\int_{{\cal D}_2}d\mu_2(y_{n-1})
\,\Delta_{n+1}({\bf y}_n,x) \prod_{j=0}^{n-1}Q_j(y_j) \cr
&={1\over\prod_{j=0}^n k_j} \det\left[\matrix{
\bigl(h_j\,\delta_{j,k}\bigr)_{j,k=0,\ldots,i-1} &\bigl(0\bigr)_{j=0,\dots,i-1\atop k=i,\ldots,n}\cr
\bigl(\gamma_{j,k}\bigr)_{j=i\,\ldots,n-1\atop k=0,\ldots,i-1}
&\bigl(\gamma_{j,k}\bigr)_{j=i\,\ldots,n-1\atop k=i,\ldots,n}\cr
\bigl(Q_k(x)\bigr)_{k=0,\ldots,i-1} &\bigl(Q_k(x)\bigr)_{k=i,\ldots,n}\cr
}\right]\,. &(3.37)\cr
}$$
The {\it Laplace expansion}
\note{See, e.g., [18] section 2.5.1.}
of this determinant according to its $i$ first rows,
and equation (3.13) yield the integral representation:
$$\eqalignno{
\widehat P_{i;n}(x) &={\prod_{j=0}^{n-1} k_j\over Z_{i;n-1} \prod_{j=0}^{i-1}h_j} \cr
\times\,\int_{\cal D}d\mu(y_0)\cdots\int_{\cal D}d\mu(y_{i-1})
&\int_{{\cal D}_2}d\mu_2(y_i)\cdots\int_{{\cal D}_2}d\mu_2(y_{n-1})
\,\Delta_{n+1}({\bf y}_n,x) \prod_{j=0}^{n-1}Q_j(y_j)\,. &(3.38)\cr
}$$
For $i=0$, this integral representation reads
$$\widehat P_{0;n}(x)={\prod_{j=0}^{n-1} k_j\over Z_{0;n-1}}
\int_{{\cal D}_2}d\mu_2(y_0)\cdots\int_{{\cal D}_2}d\mu_2(y_{n-1})
\,\Delta_{n+1}({\bf y}_n,x) \prod_{j=0}^{n-1}Q_j(y_j)
\,. \eqno(3.39)$$
Now, with the same trick as in (i) above, the measure being symmetric and $\Delta_{n+1}({\bf y}_n,x)$
being anti-symmetric in $y_0,\ldots,y_{n-1}$, one can replace $\prod_{j=0}^{n-1}Q_j(y_j)$ by
the anti-symmetric function $\det[Q_k(y_j)]_{j,k=0,\ldots,n-1}$ provided one divides the result by $n!$.
Then, from equations (3.30), (3.35) and (3.36), one recovers the known integral representation for standard
monic orthogonal polynomials of degree $n$ for the measure $\{{\cal D}_2,d\mu_2\}$
\note{See, e.g., [12] (2.2.7) and (2.2.10) where, the polynomials are normalized
according to equation (A.11), or [14] 10.3 (3)--(5).},
$$\widehat P_{0;n}(x)={1\over n!\,\det\bigl[c_{2;j+k}\bigr]_{j,k=i,\ldots,n-1}}
\int_{{\cal D}_2}d\mu_2(y_0)\cdots\int_{{\cal D}_2}d\mu_2(y_{n-1})
\,\Delta_n({\bf y}_n)^2 \prod_{j=0}^{n-1}(x-y_j)\,. \eqno(3.40)$$

\medskip
\noindent{\it 3.3.3. Domains and measures symmetric with respect to the origin.}
Let the domains $\cal D$ and ${\cal D}_2$, and the measures $d\mu$ and $d\mu_2$
be symmetric with respect to the origin, e.g., ${\cal D}=[a,b]$ with $a=-b$ and $d\mu(x)=w(x)dx$
with an even weight function $w(-x)=w(x)$, and similar relations for the measure $\{{\cal D}_2,d\mu_2\}$.
Then, as the standard orthogonal polynomials $Q_n,\,n=0,1,\ldots$ (see appendix A), the polynomials
$P_{i;n},\,n=i,i+1,\ldots$ are even or odd according as $n$ is even or odd,
$$P_{i;n}(-x)=(-1)^n P_{i;n}(x)\,. \eqno(3.41)$$
Indeed, $P_{i;n}(-x),\,n=i,i+1,\ldots$ satisfies the same orthogonality relations (3.6) as $P_{i;n}(x)$,
and since there are unique polynomials satisfying these orthogonality relation and the normalization
condition (3.8), comparing the coefficient of $x^n$ yields equation (3.41).
Actually, the even and odd polynomials separate from each other and equations (3.13)--(3.15) take the form
for $0\le i\le n$
\note{It is recalled that here, and in what follows, if $i=0$, the expressions with the
index $2i-1$ do not occur.},
$$\eqalignno{
\widehat P_{2i-1;2n}=\widehat P_{2i;2n}
&\phantom{:}=k_{2n}^{-1}\,{1\over Z_{i;n-1}^{({\rm e})}} \det\left[\matrix{
\bigl(\gamma_{2j,2k}\bigr) \cr
\noalign{\smallskip}
\bigl(Q_{2k}\bigr) \cr
}\right]_{j=i,\ldots,n-1\atop k=i,\ldots,n\hfill} &(3.42)\cr
\noalign{\smallskip}
Z_{i;n}^{({\rm e})} &:=det\bigl[\gamma_{2j,2k}\bigr]_{j,k=i,\ldots,n}\qquad Z_{i;i-1}^{({\rm e})}:=1 &(3.43)\cr
\noalign{\smallskip}
\widehat H_{2i-1;2n}=\widehat H_{2i;2n} &\phantom{:}=k_{2n}^{-2}
\,{Z_{i;n}^{({\rm e})}\over Z_{i;n-1}^{({\rm e})}} &(3.44)\cr
\noalign{\medskip}
\widehat P_{2i;2n+1}=\widehat P_{2i+1;2n+1}
&\phantom{:}=k_{2n+1}^{-1}\,{1\over\,Z_{i;n-1}^{({\rm o})}} \det\left[\matrix{
\bigl(\gamma_{2j+1,2k+1}\bigr) \cr
\noalign{\smallskip}
\bigl(Q_{2k+1}\bigr) \cr
}\right]_{j=i,\ldots,n-1\atop k=i,\ldots,n\hfill} &(3.45)\cr
\noalign{\smallskip}
Z_{i;n}^{({\rm o})} &:=det\bigl[\gamma_{2j+1,2k+1}\bigr]_{j,k=i,\ldots,n}\qquad Z_{i;i-1}^{({\rm o})}:=1 &(3.46)\cr
\noalign{\smallskip}
\widehat H_{2i;2n+1}=\widehat H_{2i+1;2n+1} &\phantom{:}=k_{2n+1}^{-2}
\,{Z_{i;n}^{({\rm o})}\over Z_{i;n-1}^{({\rm o})}}\,. &(3.47)\cr
}$$
As for the standard orthogonal polynomials (see equations (A.17)--(A.28)), these relations can be obtained in
two ways:
(1) it can be checked, as for equation (2.6), that these monic polynomials
satisfy the orthogonality conditions (3.6) for $(\,,\,)$ and $(\,,\,)_2$ which define them uniquely;
(2) taking advantage of the checkerboard structure of the matrix associated with the metric tensor,
since $\gamma_{j,k}$ vanishes if $j+k$ odd, the determinants which occur in the general formulae
(3.13)--(3.17) can be evaluated from the lemma given in appendix A (see equations (A.13)--(A.15)).
As an example, let us compute $\widehat P_{2i;2n}$, one has
$$\eqalignno{
\widehat P_{2i;2n} &=k_{2n}^{-1}\,{1\over Z_{2i;2n-1}} \det\left[\matrix{
\bigl(\gamma_{j,k}\bigr) \cr
\noalign{\smallskip}
\bigl(Q_k\bigr) \cr
}\right]_{j=2i,\ldots,2n-1\atop k=2i,\ldots,2n\hfill} \cr
&=k_{2n}^{-1}\,{1\over Z_{2i;2n-1}} \det\left[\matrix{
\bigl(\gamma_{2j,2k}\bigr) \cr
\noalign{\smallskip}
\bigl(Q_{2k}\bigr) \cr
}\right]_{j=i,\ldots,n-1\atop k=i,\ldots,n\hfill} \det\bigl[\gamma_{2j+1,2k+1}\bigr]_{j,k=i,\ldots,n-1} &(3.48)\cr
}$$
which, as expected, does not depend on the odd polynomials $Q_{2j+1},\,j=i,\ldots,n-1$,
and $Z_{2i;2n-1}$ reads
$$Z_{2i;2n-1}=\det\bigl[\gamma_{j,k}\bigr]_{j,k=2i,\ldots,2n-1}
=\det\bigl[\gamma_{2j,2k}\bigr]_{j,k=i,\ldots,n-1} \det\bigl[\gamma_{2j+1,2k+1}\bigr]_{j,k=i,\ldots,n-1}\,.
\eqno(3.49)$$
Thereby, one gets the expressions (3.42) and (3.43) of $\widehat P_{2i;2n}$ and $Z_{i;n}^{({\rm e})}$, respectively.

It follows from equations (3.42) and (3.45) that
$${\rm if}\ (-1)^{i+n}=1\quad{\rm then}\quad \widehat P_{i-1;n}=\widehat P_{i;n}\quad0\le i-1\le n\,.
\eqno(3.50)$$
In other words, in the $i,n$-plane, two neighboring points having the same ordinate $n$ are associated with
equal SBO polynomials if both coordinates $i$ and $n$ of the right-hand point have the same parity.
As a special case, the points immediately on the left of the boundary $i=n\ge1$ are also associated with
known SBO polynomials, since from equation (3.13) one has
$$\widehat P_{i-1;i}=\widehat P_{i;i}=\widehat Q_i\quad i\ge1\,. \eqno(3.51)$$
Therefore, in addition to $\widehat P_{0;0}=1$, one has $\widehat P_{0;1}=\widehat P_{1;1}=x$.

Note that if $i=1$, setting $i=0$ in equation (3.45) yields $\widehat P_{1;2n+1}$ for $n=0,1,\ldots$,
$$\widehat P_{1;2n+1}=\widehat Q_{2;2n+1}=\widehat P_{0;2n+1} \eqno(3.52)$$
as it should be from equations (3.18) and (3.50).
Actually, the identity between $\widehat P_{1;2n+1}$ and $\widehat Q_{2;2n+1}$ follows readily from parity
arguments.
Indeed, the odd standard orthogonal polynomials with respect to $(\,,\,)_2$, i.e.
$\widehat Q_{2;2n+1},\,n=0,1,\ldots$,
are orthogonal with any even polynomial with respect to any inner product defined with an even measure.
Thus, they are orthogonal to $x^0$ with respect to $(\,,\,)$ and orthogonal to $\widehat P_{1;2n}$
with respect to $(\,,\,)_2$.

\medskip
\noindent{\it 3.3.4. Normalization.}
As for the standard orthogonal polynomials (see appendix A) three particular choices of normalization
can be considered with interest:

\noindent (i) $K_{i;n}=1$, then $P_{i;n}=\widehat P_{i;n}$ is a monic polynomial;

\noindent (ii) $H_{i;n}=1$, then the polynomials $P_{i;n},\,n=0,1,\ldots$ are orthonormal.
In addition, choosing $K_{i;n}/k_n>0$, from equation (3.15), one has
$${K_{i;n}\over k_n}=\Bigl({Z_{i;n-1}\over Z_{i;n}}\Bigr)^{1/2}
\qquad P_{i;n}=\bigl({Z_{i;n-1} Z_{i;n}}\bigr)^{-1/2} \det\left[\matrix{
\bigl(\gamma_{j,k}\bigr) \cr
\noalign{\smallskip}
\bigl(Q_k\bigr) \cr
}\right]_{j=i,\ldots,n-1\atop k=i,\ldots,n\hfill}
\qquad  P_{i;i}=Z_{i,i}^{-1/2} Q_i\,; \eqno(3.53)$$

\noindent (iii) $K_{i;n}=k_n Z_{i;n-1}$, then $Z$-factors in denominators remain only in equation (3.17),
in particular,
$$H_{i;n}=Z_{i;n-1} Z_{i;n}\qquad P_{i;n}=\det\left[\matrix{
\bigl(\gamma_{j,k}\bigr) \cr
\noalign{\smallskip}
\bigl(Q_k\bigr) \cr
}\right]_{j=i,\ldots,n-1\atop k=i,\ldots,n\hfill}\,. \eqno(3.54)$$

\medskip
\noindent{\it 3.3.5. Linear recurrence formula with respect to the degree $n$.}
Standard orthogonal polynomials fulfill the {\it three term linear recurrence formula}
for $n=0,1,\ldots$
\note{See, e.g., [12] theorem 3.2.1 or [14] 10.3 (7) and (8).
$A_n$ and $B_n$ are such that $Q_{n+1}-(A_n x+B_n) Q_n$ is of degree at most $n-1$,
thus, it reads $\sum_{j=0}^{n-1} \alpha_j Q_j$ where, from the orthogonality relation (A.2),
for $j=0,\ldots,n-1$,
$\alpha_j h_j=-(\widehat Q_j\,,\,A_n x Q_n)=-A_n(x Q_j\,,\,Q_n)$.
Since $x Q_j$ is of degree $j+1$, the last inner product vanishes except for $j=n-1$.
The coefficient of $x^n$ in $x Q_{n-1}$ being $k_{n-1}$,
this fixes $\alpha_{n-1}=-C_n$ as given in equation (3.55).},
$$Q_{n+1}=(A_n x+B_n) Q_n-C_n Q_{n-1}
\qquad A_n ={k_{n+1}\over k_n}\quad B_n=A_n(\widehat r_{n+1}-\widehat r_n)
\quad C_n={A_n\over A_{n-1}} {h_n\over h_{n-1}}\,. \eqno(3.55)$$
Following the same steps, with $\widehat B_{i;n}:=\widehat R_{i;n+1}-\widehat R_{i;n}$,
the polynomial $\widehat P_{i;n+1}-(x+\widehat B_{i;n})\,\widehat P_{i;n}$ is of degree at most $n-1$.
Therefore, it reads $\sum_{j=0}^{n-1} \alpha_{i;j}\,\widehat P_{i;j}$ where, from the orthogonality condition (3.6),
$$\alpha_{i;j} \widehat H_{i;j}
=\bigl(\widehat P_{i;j}\,,\,\widehat P_{i;n+1}-(x+\widehat B_{i;n}) \widehat P_{i;n}\bigr)_0
=-\bigl(\widehat P_{i;j}\,,\,x \widehat P_{i;n}\bigr)_0\,. \eqno(3.56)$$
But now, for $i\le n$, $(\widehat P_{i;j}\,,\,x \widehat P_{i;n})_0$ is not equal to
$(x \widehat P_{i;j}\,,\,\widehat P_{i;n})_0$, since the inner product $(\,,\,)_0$
is not defined by a single integral.
From equation (3.5), to evaluate $(\widehat P_{i;j}\,,\,x\,\widehat P_{i;n})_0$, one has to split
$x \widehat P_{i;n}$ into its components in ${\cal P}_i$ and in ${\cal P}_{i;N}^\bot$,
e.g., by expanding $x \widehat P_{i;n}$ on the basis polynomials $\widehat P_{i;j},\,j=0,\ldots,n+1$.
This can be done as follows using equations (3.7) and (3.55), for $i\le n$,
$$\eqalignno{
x \widehat P_{i;n} &=\sum_{m=i}^n x Q_m \widehat A_{i;m,n}
=\sum_{m=i}^n A_m^{-1} (Q_{m+1}-B_m Q_m+C_m Q_{m-1}) \widehat A_{i;m,n} \cr
&=\beta_{i;i-1} Q_{i-1}+\sum_{m=i}^{n+1} \beta_{i;m} Q_m
=\beta_{i;i-1} Q_{i-1}+\sum_{m=i}^{n+1} \eta_{i;m} \widehat P_{i;m} &(3.57)\cr
}$$
where, $\beta_{i;i-1}$, $\beta_{i;m}$ and $\eta_{i;m}$ for $m=i,\ldots,n+1$
can be easily written in terms of $\widehat A_{i;j,k}$, $\widehat B_{i;j,k}$, $A_j$, $B_j$ and $C_j$
for $i\le j\le k\le n+1$.
In the subspace ${\cal P}_i$, the expansion of $Q_{i-1}$ in terms of $\widehat P_{i;j},\,j=0,\ldots,i-1$
depends on the inner product $(\,,\,)_1$
\note{If $(\,,\,)_1=(\,,\,)$, then $\widehat P_{i;j}=\widehat Q_j,\,j=0,\ldots,i-1$ and
$\alpha_{i;j}$ vanishes for $j=0,\ldots,i-2$.}.
In the generic case, all the coefficients $\alpha_{i;j}=-\eta_{i;j},\,j=i,\ldots,n-1$ are nonzero,
while for the standard polynomials, only the equivalent coefficient for $j=n-1$ does not vanish.
Thus, there is no generalization of the standard three term linear recurrence formula involving a constant number
of terms.

\smallskip
However, five term linear recurrence formulae are derived in a forthcoming paper II for
special cases associated with Hermite and Laguerre polynomials.
These relations, as also some differentiation formulae, are based on differential equations satisfied by
the weight functions $w$ and $w_2$ we consider.

\medskip
\noindent{\it 3.3.6. Linear recurrence formula with respect to both $i$ and $n$.}
Equations (3.7), (3.16) and (3.17) allow to relate polynomials $\widehat P_{i;n}$ corresponding to
different values of $i$.
With $0\le i\le j\le n$, inverting the finite sums over $m$ and $\ell$, one has
$$\widehat P_{j;n}=\sum_{m=j}^n \Bigl(\sum_{\ell=i}^m \widehat P_{i;\ell} \widehat B_{i;\ell,m}\Bigr)
\widehat A_{j;m,n}=\sum_{\ell=i}^n \widehat P_{i;\ell} \widehat C_{i,j;\ell,n}
\qquad\widehat C_{i,j;\ell,n}:=\sum_{m=\sup(j,\ell)}^n \widehat B_{i;\ell,m} \widehat A_{j;m,n} \eqno(3.58)$$
where $\widehat C_{i,i;\ell,n}=\delta_{\ell,n}$.
Furthermore, since $\widehat P_{i;n}$ and $\widehat P_{j;n}$ are monic polynomials, $\widehat C_{i,j;n,n}=1$.
Setting $j=i+1$, and provided one is able to compute the coefficients $\widehat C_{i,i+1;\ell,n}$,
the linear recurrence formula (3.58) allows to determine step by step $\widehat P_{i;n}$ for $0\le i\le n$,
starting with $\widehat P_{0;n}$ given by equation (3.18).
This is done explicitly in a forthcoming paper II for special cases associated with the Hermite
and Laguerre polynomials.

\medskip
\noindent{\it 3.3.7. Properties of the zeros.}
Several properties of the zeros of standard orthogonal polynomials have been known for a long time:
e.g.,
{\sl all the zeros are real, simple and located in the support ${\cal D}:=[a,b]$ of the measure} and also,
{\sl the zeros of polynomials with consecutive degrees separate each other}
\note{See, e.g., [12] theorems 3.3.1 and 3.3.2.}.
We saw previously that the SBO polynomial $P_{i;n}$ coincides with standard
orthogonal polynomials in several special cases, i.e. from equations (3.13) and (3.18),
$$\widehat P_{n;n}=\widehat Q_n\qquad\widehat P_{0;n}=\widehat Q_{2;n}\quad n=0,1,\ldots \eqno(3.59)$$
and furthermore, if the domains ${\cal D}$ and ${\cal D}_2$, and the measures $d\mu$ and $d\mu_2$ are symmetric
with respect to the origin, from equations (3.51) and (3.50),
$$\widehat P_{n-1;n}=\widehat Q_n\qquad\widehat P_{1;2n-1}=\widehat Q_{2;2n-1}\quad n=1,2,\ldots\,. \eqno(3.60)$$

\smallskip
Now, what about the zeros of $P_{i;n}$ in the remaining generic case $0<i<n$?
Starting the proof as for the standard orthogonal polynomials (see footnote 42), one has from equation (3.6),
$$(P_{i;n}\,,\,P_{i;0})_0=(P_{i;n}\,,\,P_{i;0})\propto\int_{\cal D} d\mu(x) P_{i;n}(x)=0\,. \eqno(3.61)$$
Hence, the measure being positive, there exists at least one value of $x$ in ${\cal D}:=[a,b]$ where $P_{i;n}$
changes sign.
If $P_{i;n}$ changed its sign in ${\cal D}$ at $m<i$ points $x_1,\ldots,x_m$,
the polynomial $R_m:=\prod_{j=1}^m (x-x_j)$ of degree $m$ would be such that,
$$(P_{i;n}\,,\,R_m)_0=(P_{i;n}\,,\,R_m)=\int_{\cal D} d\mu(x) P_{i;n}(x) R_m(x)=0\,. \eqno(3.62)$$
This contradicts the fact that $P_{i;n}(x) R_m(x)$ would have a constant sign throughout ${\cal D}$.
Therefore:
{\sl for  $0\le i\le n$, $P_{i;n}$ has at least $m\ge i$ distinct real zeros of odd order in ${\cal D}$}.
Also, $m\le n$, since $P_{i;n}$ is of exact degree $n$.
Since the coefficients of the polynomials considered are real, if $x$ is a zero, so is its conjugate complex.
Therefore, it follows that
{\sl $P_{n-1;n},\,n=1,2,\ldots$ has $n$ real and simple zeros in ${\cal D}$},
generalizing thereby the property following from equation (3.60) only for symmetric measures.
On the other hand, the usual argument above cannot be applied when $i\le m\le n$ since,
although $(P_{i;n}\,,\,R_m)_0=0$, this inner product is no longer defined by a single integral.

\smallskip
Concerning the relative positions of the zeros of SBO polynomials with a given $i$ and consecutive
degrees, we have not been able to prove a general property.
Let us only recall that the proof of the interleaving of the zeros for standard orthogonal polynomials
is a consequence of the {\it Christoffel-Darboux formula}
\note{See, e.g., [12] theorem 3.2.2 or [14] 10.3 (10) and (11).},
which itself follows from the three term recurrence formula.
It has already been emphasized in section 3.3.5 that such a recurrence formula no longer holds for
SBO polynomials.

\medskip
\bigskip
\noindent{\elevenrmb 4. Conclusion}

\bigskip
\noindent In order to take into account the particle-number conservation within the density functional theory,
B Giraud {\it et al} [1--3] considered a new sets of real `constrained polynomials'.
These polynomials satisfy a constraint of vanishing average with a positive measure
and are orthogonal with respect to a inner product defined by a second positive measure.
The linear constraint considered can be viewed as an orthogonality relation with a constant polynomial,
with respect to another inner product defined by the first measure.
This allows us to recast the determination of these polynomials into a more general problem of
finding particular orthogonal bases in an Euclidean vector space endowed with distinct Euclidean inner products.

Recalling basic properties of linear algebra, it is shown that, given an Euclidean vector space ${\cal E}$
with the inner product $(\,,\,)$, and any $i$-dimensional subspace  ${\cal E}_i$,
there exists a unique complementary and orthogonal subspace ${\cal E}_i^\bot$ of codimension $i$.
Endowing each of these subspaces with different Euclidean inner products $(\,,\,)_1$ and $(\,,\,)_2$,
respectively, yields to define  ${\cal E}_i$ and ${\cal E}_i^\bot$ as BO subspaces.
This induces a new Euclidean inner product $(\,,\,)_0$ on ${\cal E}$, which coincides with $(\,,\,)_1$
on ${\cal E}_i$ and with $(\,,\,)_2$ on ${\cal E}_i^\bot$, and vanishes otherwise.
An orthogonal basis with respect to $(\,,\,)_0$ is called a BO basis, providing a general frame
to study particular constrained polynomials.
We give a general strategy to determine such a basis using a two step G-SO procedure.
Let $\{e_0,\ldots,e_{i-1}\}$ be any basis of ${\cal E}_i$.
A first step determines ${\cal E}_i^\bot$ by G-SO of any basis $\{e_0,\ldots,e_{i-1},e_i,\ldots\}$
of ${\cal E}$, with respect to the inner product $(\,,\,)$.
This provides an orthogonal basis $\{E_0,E_1,\ldots\}$ of ${\cal E}$ such that $\{E_i,E_{i+1},\ldots\}$
is a basis of ${\cal E}_i^\bot$.
In a second step, a G-SO of $\{E_i,E_{i+1},\ldots\}$ with respect to the inner product $(\,,\,)_2$
provides an orthogonal basis $\{\phi_i,\phi_{i+1},\ldots\}$ of ${\cal E}_i^\bot$.
Such a basis depends on ${\cal E}_i$ and on the two inner products $(\,,\,)$ and $(\,,\,)_2$.
(The standard problem of getting an orthogonal basis $\{\phi_0,\ldots,\phi_{i-1}\}$ of ${\cal E}_i$
with respect to $(\,,\,)_1$ brings nothing new.)
All the relevant quantities in the G-SO (e.g., connection coefficients between bases) are expressed in terms
of determinants with metric tensor components as entries, except possibly for the last row.
Furthermore, the G-SO is a step by step procedure well-suited for numerical calculation.
The importance of the non-degenerate and/or the definite character of the inner
products considered is underlined.
Since different inner products are considered simultaneously, the Dirac notation must be handled with care,
in particular to express the orthogonal projectors onto ${\cal E}_i$ and ${\cal E}_i^\bot$.
All these considerations can be extended to Hermitian vector spaces and to Hilbert spaces.

These general results apply readily to the special case of vector spaces of real polynomials with Euclidean
inner products defined with positive measures.
The generalization of the problem of constrained polynomials is as follows.
Let $(\,,\,)$ be an Euclidean inner product and $\{e_0,\ldots,e_{i-1}\}$ be any $i$ linearly independent
polynomials defining the $i$-dimensional subspace ${\cal E}_i$.
The polynomials $P$, satisfying the $i$ constraints $(e_j\,,\,P)=0,\,j=0,\ldots,i-1$, belong to the unique subspace
${\cal E}_i^\bot$ of codimension $i$ which is complementary and orthogonal to ${\cal E}_i$ with respect
to $(\,,\,)$.
An orthogonal basis of ${\cal E}_i^\bot$ with respect to a distinct Euclidean inner product $(\,,\,)_2$
defines a BO basis.
Now, the structure of polynomial algebra with the multiplication law $x^j x^k=x^{j+k}$,
and the definition of the inner products in terms of integrals provide additional properties.

We make a systematic study of these properties in the special case where ${\cal E}_i$ is the subspace
${\cal P}_i$ of real polynomials of degree less than $i$.
Then, the first step to define ${\cal P}_i^\bot$ coincides with the determination of the standard
orthogonal polynomials $Q_0,Q_1,\ldots$ with respect to the inner product $(\,,\,)$, $Q_n$ being of exact degree
$n$.
Requiring that the BO basis polynomials $P_{i;n}:=\phi_n,\,n=i,i+1,\ldots$ of ${\cal P}_i^\bot$,
be of exact degree $n$, defines them uniquely apart from a constant multiplicative factor for each polynomial.
We call them the SBO polynomials.
Emphasizing the similarities to and the differences from the standard orthogonal polynomials
(basically due to the fact that inner product $(\,,\,)_0$ is no longer defined by a single integral),
we investigate the projection operators, the integral representations, the special case of symmetric measures
with respect to the origin, the normalizations, the recurrence formulae with respect to the degree $n$
and with respect to both $i$ and $n$ and finally the zeros.
For physical applications, it is useful to consider the special case where the inner product $(\,,\,)$ is
defined by the classical measure corresponding to a classical orthogonal polynomial set.
This will be done in a forthcoming paper II for Hermite and Laguerre polynomials.

Applying the general formulae, a similar calculation can be done with a choice of the subspace
of polynomials ${\cal E}_i$ different from ${\cal P}_i$,
Then, the BO polynomials are not what we call the {\it standard} ones,
and a priori, a new similar study starting from the general formulae (2.4)--(2.10) is required.
Note that if the measures considered are symmetric with respect to the origin (see section 3.3.3),
and if ${\cal E}_i$ is a subspace of polynomials with a given parity, then, due to trivial parity arguments,
the monic BO polynomials of the opposite parity coincide with the monic standard BO polynomials $\widehat Q_{2;n}$
of the same degree.
For example, for a given $m\ge0$, let ${\cal E}_1$ be spanned by $\{x^{2m+1}\}$ associated with $i=1$
constraint of zero odd $(2m+1)$th moment.
Then, the even monic BO polynomials read $\widehat P_{2n}=\widehat Q_{2;2n},\,n=0,1,\ldots$.
The odd BO polynomials require a new determination, except for $m=0$, where,
the even and odd polynomials being decoupled and $x$ being the lowest degree odd polynomial,
the problem reduces to the same study as for the odd SBO polynomials in the case $i=2$.
Thus, one has $\widehat P_{2n+1}=\widehat P_{2;2n+1},\,n=1,2,\ldots$.
Such a case has already been considered in [1]
\note{The `Hermite polynomials constrained by a zero momentum', considered in [1] section 3 (17),
with the weight functions $w:=e^{-{1\over2}x^2}$ and $w_2:=e^{-x^2}$, {\it is not the case} $i=2$.
Indeed, the one-dimensional subspace ${\cal E}_1$ considered there, and spanned by $\{x^1\}$,
is not ${\cal P}_2$ spanned by $\{x^0,x^1\}$.
Nevertheless, it follows from above that the monic BO polynomials are:
$\widehat P_{2n}=\widehat Q_{2;2n},\,n=0,1,\ldots$ and $\widehat P_{2n+1}=\widehat P_{2;2n+1},\,n=1,2,\ldots$.
However, this simplification would no longer hold for measures not symmetric with respect to the origin and/or,
e.g., Hermite polynomials constrained by zero first and second moments,
i.e. now with ${\cal E}_2$ spanned by $\{x^1,x^2\}$).
Then, the BO polynomials would not be the {\it standard} ones.
}.

More generally, instead of constraints all associated with the same inner product $(\,,\,)$,
one may think to use a similar approach to take into accounts constraints associated with different inner
products.
As a first example of generalization, does there exist vectors $v$ (e.g. polynomials) such that, $(e_1\,,\,v)_1=0$
and $(e_2\,,\,v)_2=0$
(i.e. for two given distinct vectors $e_1$ and $e_2$, and two given distinct inner products $(\,,\,)_1$ and $(\,,\,)_2$)?
We show in section 2.3.1 that this generalization of BO subspaces to more than two subspaces may have no
solution, a unique solution or an infinite number of solutions according to the inner products involved
between pairs of subspaces.
As a second example of generalization, does there exist vectors $v$ such that $(e_1\,,\,v)=0$ and $(e_1\,,\,v)'=0$
(i.e. for one given vector $e_1$, and two given distinct inner products $(\,,\,)$ and $(\,,\,)'$)?
We argue in section 2.3.2 that, for this example, $v$ belong to a subspace of dimension 1 or 0 according to
the inner products considered.
One of the referees brought our attention to a similar study on {\it multiple orthogonal polynomials} or
{\it Hermite-Pade polynomials} [21], e.g., with $0\le m\le n$, polynomials $P_{m,n}$ of degree $n$ orthogonal
with respect to a weight function $w_1$ to $x^j,\,j=0,\ldots,m$ and orthogonal with respect to another weight
function $w_2$ to $x^j,\,j=0,\ldots,n-m-2$.
Assuming the weigt functions are {\it classical}, i.e. solution of Pearson's differential equation
$(\phi w)'+\psi w=0$, with polynomial coefficients $\phi$ and $\psi$,
these polynomials have many new applications and nontrivial
generalization of the theory of standard orthogonal polynomials.

\medskip
\bigskip
\noindent{\elevenrmb Acknowledgments}

\bigskip
\noindent B Giraud drew my attention to constrained orthogonal polynomials.
I had many helpful and friendly discussions about polynomials with M L Mehta at the beginning of this work.
B Eynard suggested the expression in terms of a determinant for constrained polynomials.
I also had some stimulating discussions with P Moussa.
It is a pleasure to thank all of them, and especially M L Mehta.
I am grateful to D Kosower who helped me to improve the English language.
Finally, we thank referees for pointing us a similar work on multiple orthogonal polynomials
and also providing us with some additional references on basic properties on functional analysis and
orthogonal polynomials.

\bigskip
\bigskip
\noindent{\elevenrmb Appendix A. Standard orthogonal polynomials}

\medskip
\noindent As considered in section 3.1, let ${\cal P}_N$ be the vector space over the real field of real
polynomials of degree less than $N$ with the Euclidean inner product $(\,,\,)$ defined with the measure
$\{{\cal D},d\mu\}$.
With respect to the monomial basis $\{x^0,\ldots,x^{N-1}\}$, the matrix of metric tensor components is a
{\it persymmetric matrix} known as the {\it moment} or {\it Hankel matrix}
\note{See, e.g., [18] section 7.4.
This basic property follows from the polynomial algebra, see footnote 29.},
which entries are the moments of the measure defined by equation (3.1)
\note{As noted in [22] corollary 6.3.2, one can consider as well
$g_{j,k}:=(\varepsilon_j\,,\,x^k)$, where, $\varepsilon_j$ is any polynomial of exact degree $j$,
in order to make easier the forthcoming calculations.},
$$g_{j,k}=(x^j\,,\,x^k)
=\int_{\cal D}d\mu(x)\,x^{j+k}=c_{j+k}\quad j,k=0,\ldots,N-1\,. \eqno(A.1)$$

\smallskip
\noindent{\it A.1. Relations between standard orthogonal polynomials and monomials}

\smallskip
\noindent For $0\le n\le N-1$ and given arbitrary nonzero finite $k_j:=b_{j,j}^{-1},\,j=0,\ldots,n$,
according to equations (2.4)--(2.10), the G-SO of
$\{x^0,\ldots,x^n\}$ defines {\it uniquely} the standard orthogonal polynomials $Q_j,\,j=0,\ldots,n$,
such that $Q_j$ is of degree $j$ with $k_j$ as coefficient of $x^j$,
$$\eqalignno{
(Q_j\,,\,Q_k) &=h_j \delta_{j,k}\quad j,k=0,\ldots,n &(A.2)\cr
\noalign{\smallskip}
Q_n &=\sum_{m=0}^n x^m a_{m,n}:=k_n\,\widehat Q_n
\qquad\widehat Q_n:=x^n+\widehat r_n x^{n-1}+\widehat s_n x^{n-2}+{\rm O}(x^{n-3}) &(A.3)\cr
\noalign{\smallskip}
x^n &\phantom{:}=\sum_{m=0}^n Q_m b_{m,n}\qquad b_{n,n}=a_{n,n}^{-1}=k_n^{-1} &(A.4)\cr
}$$
where $\widehat Q_n$ denotes the monic polynomial
\note{The coefficient $\widehat r_n$ is denoted $r_n$ by [14], see section 10.3 p 158,
where the coefficient $k'_n:=k_n\,r_n$ of $x^{n-1}$ in $Q_n$ is also introduced.}.
One has for $0\le m\le n\le N-1$,
$$\eqalignno{
Z_n &\phantom{:}=det\bigl[c_{j+k}\bigr]_{j,k=0,\ldots,n}\qquad Z_{-1}:=1 &(A.5)\cr
\noalign{\smallskip}
\widehat Q_n &\phantom{:}={1\over Z_{n-1}} \det\left[\matrix{
\bigl(c_{j+k}\bigr) \cr
\noalign{\smallskip}
\bigl(x^k\bigr) \cr
}\right]_{j=0,\ldots,n-1\atop k=0,\ldots,n\hfill}\qquad \widehat Q_0=1 &(A.6)\cr
\noalign{\smallskip}
h_n &:=k_n^2\,\widehat h_n\qquad\widehat h_n={Z_n\over Z_{n-1}} &(A.7)\cr
\noalign{\smallskip}
a_{m,n} &:=k_n\,\widehat a_{m,n}\qquad\widehat a_{m,n}={1\over Z_{n-1}} \det\left[\matrix{
\bigl(c_{j+k}\bigr) \cr
\noalign{\smallskip}
\bigl(\delta_{m,k}\bigr) \cr
}\right]_{j=0,\ldots,n-1\atop k=0,\ldots,n\hfill} &(A.8)\cr
\noalign{\smallskip}
\widehat r_n &\phantom{:}=\widehat a_{n-1,n}\qquad\widehat s_n=\widehat a_{n-2,n}
\qquad\widehat Q_n(0)=\widehat a_{0,n} &(A.9)\cr
\noalign{\smallskip}
b_{m,n} &:=k_m^{-1}\,\widehat b_{m,n}\qquad\widehat b_{m,n}={1\over Z_m} \det\left[\matrix{
\bigl(c_{j+k}\bigr) \cr
\noalign{\smallskip}
\bigl(c_{n+k}\bigr) \cr
}\right]_{j=0,\ldots,m-1\atop k=0,\ldots,m\hfill}\,. &(A.10)
}$$

Apart from the usual choice of normalization for the classical orthogonal polynomials
\note{See, e.g., [14] 10.13 (4), 10.12 (2), 10.9 (8) (for $\lambda\ne0$) and 10.8 (5) for the choice of $k_n$
for Hermite, Laguerre, Gegenbauer and Jacobi polynomials, respectively.},
three particular normalizations are commonly used:

\noindent (i) $k_n=1$, then $Q_n=\widehat Q_n$ is a monic polynomial;

\noindent (ii) $h_n=1$, then the polynomials $Q_n,\,n=0,1,\ldots$ are orthonormal.
In addition, choosing $k_n>0$, from equation (A.7), one has
\note{See, e.g., [12] (2.2.6), (2.2.7) and (2.2.15) where, $D_n$ denotes $Z_n$.},
$$k_n=\Bigl({Z_{n-1}\over Z_n}\Bigr)^{1/2}\qquad Q_n=\bigl(Z_{n-1} Z_n\bigr)^{-1/2} \det\left[\matrix{
\bigl(c_{j+k}\bigr) \cr
\noalign{\smallskip}
\bigl(x^k\bigr) \cr
}\right]_{j=0,\ldots,n-1\atop k=0,\ldots,n\hfill}\qquad Q_0=Z_0^{-1/2}\,; \eqno(A.11)$$

\noindent (iii) $k_n=Z_{n-1}$, then $Z$-factors in denominators remain only in equation (A.10), in particular,
$$h_n=Z_{n-1} Z_n\qquad Q_n=\det\left[\matrix{
\bigl(c_{j+k}\bigr) \cr
\noalign{\smallskip}
\bigl(x^k\bigr) \cr
}\right]_{j=0,\ldots,n-1\atop k=0,\ldots,n\hfill}\qquad Q_0=1\,. \eqno(A.12)$$

\smallskip
\noindent{\it A.2. Domain and measure symmetric with respect to the origin}

\smallskip
\noindent If, for example, ${\cal D}=[a,b]$
with $a=-b$ and $d\mu(x)=w(x)dx$ with an even weight function $w(-x)=w(x)$, the polynomials $Q_n,\,n=0,1,\ldots$
are even or odd according as $n$ is even or odd.
Indeed, $Q_n(-x)$ satisfies the same orthogonality relations (A.2) as $Q_n(x)$,
and since there are unique polynomials satisfying  these orthogonality relations with the normalization (A.3),
comparing the coefficient of $x^n$ yields $Q_n(-x)=(-1)^n\,Q_n(x)$.
Actually, the even and odd polynomials separate from each other:
$c_{j+k}$ vanishes for $j+k$ odd and the determinants which occur in the general formulae (A.5)--(A.10)
have a checkerboard structure, except possibly for the last row.
These determinants can be evaluated with the following lemma:

\smallskip
\noindent{\bf Lemma A}.
{\sl
\noindent Let ${\bf A}:=(A_{j,k})_{j,k=0,\ldots,n-1}$ be an $n\times n$ matrix with the checkerboard structure
$A_{j,k}=0$ if $j+k$ {\it odd}, except possibly for the last row $j=n-1$.
Then, one has
$$\det{\bf A}=\det{\bf B} \det{\bf C} \eqno(A.13)$$
where, for $m=1,2,\ldots$,
$$\eqalignno{
n &:=2m\qquad{\bf B}:=\bigl(B_{j,k}:=A_{2j,2k}\bigr)_{j,k=0,\ldots,m-1}
\qquad {\bf C}:=\bigl(C_{j,k}:=A_{2j+1,2k+1}\bigr)_{j,k=0,\ldots,m-1} &(A.14)\cr
n &:=2m+1\qquad{\bf B}:=\bigl(B_{j,k}:=A_{2j,2k}\bigr)_{j,k=0,\ldots,m}
\qquad {\bf C}:=\bigl(C_{j,k}:=A_{2j+1,2k+1}\bigr)_{j,k=0,\ldots,m-1}\,. &(A.15)\cr
}$$
The relation above is still true for $n=1$, setting $\det{\bf C}:=1$ as a null determinant.
}

\smallskip
For $n$ even, the proof is as follows.
Permuting rows and columns of ${\bf A}$ in order to collect the zero elements into two blocks, one gets
$$\det{\bf A}=\det\pmatrix{
{\ \bf B}\hfill &\bigl(0\bigr)_{j,k=0,\ldots,m-1}\hfill\cr
\noalign{\smallskip}
\bigl(0\bigr)_{j=0,\ldots,m-2\atop k=0,\ldots,m-1}\hfill
&\bigl(A_{2j+1,2k+1}\bigr)_{j=0,\ldots,m-2\atop k=0,\ldots,m-1}\hfill\cr
\noalign{\smallskip}
\bigl(A_{n-1,2k}\bigr)_{k=0,\ldots,m-1}\hfill &\bigl(A_{n-1,2k+1}\bigr)_{k=0,\ldots,m-1}\hfill\cr
}\,. \eqno(A.16)$$
The Laplace expansion (see footnote 37) of this determinant according to its first $m$ rows yields equation (A.14)
(the result does not depend on $A_{n-1,2k},\,k=0,\ldots,m-1$).
For $n$ odd, the proof is similar (then, the result does not depend on $A_{n-1,2k+1},\,k=0,\ldots,m-1$).

Now, this lemma yields for $n=0,1,\ldots$,
$$\eqalignno{
Z_n^{({\rm e})} &:=det\bigl[c_{2j+2k}\bigr]_{j,k=0,\ldots,n}\qquad Z_{-1}^{({\rm e})}:=1 &(A.17)\cr
\noalign{\smallskip}
\widehat Q_{2n} &\phantom{:}=\sum_{m=0}^n \widehat a_{2m,2n} x^{2m}
={1\over Z_{n-1}^{({\rm e})}} \det\left[\matrix{
\bigl(c_{2j+2k}\bigr) \cr
\noalign{\smallskip}
\bigl(x^{2k}\bigr) \cr
}\right]_{j=0,\ldots,n-1\atop k=0,\ldots,n\hfill}\qquad\widehat Q_0=1 &(A.18)\cr
\noalign{\smallskip}
x^{2n} &\phantom{:}=\sum_{m=0}^n \widehat Q_{2m} \widehat b_{2m,2n} &(A.19)\cr
\noalign{\smallskip}
\widehat h_{2n} &\phantom{:}={Z_n^{({\rm e})}\over Z_{n-1}^{({\rm e})}} &(A.20)\cr
\noalign{\smallskip}
\widehat a_{2m,2n} &\phantom{:}={1\over Z_{n-1}^{({\rm e})}} \det\left[\matrix{
\bigl(c_{2j+2k}\bigr) \cr
\noalign{\smallskip}
\bigl(\delta_{m,k}\bigr) \cr
}\right]_{j=0,\ldots,n-1\atop k=0,\ldots,n\hfill} &(A.21)\cr
\noalign{\smallskip}
\widehat b_{2m,2n} &\phantom{:}={1\over Z_m^{({\rm e})}} \det\left[\matrix{
\bigl(c_{2j+2k}\bigr) \cr
\noalign{\smallskip}
\bigl(c_{2n+2k}\bigr) \cr
}\right]_{j=0,\ldots,m-1\atop k=0,\ldots,m\hfill} &(A.22)\cr
\noalign{\medskip}
Z_n^{({\rm o})} &:=det\bigl[c_{2j+2k+2}\bigr]_{j,k=0,\ldots,n}\qquad Z_{-1}^{({\rm o})}:=1 &(A.23)\cr
\noalign{\smallskip}
\widehat Q_{2n+1} &\phantom{:}=\sum_{m=0}^n \widehat a_{2m+1,2n+1} x^{2m+1}
={1\over Z_{n-1}^{({\rm o})}} \det\left[\matrix{
\bigl(c_{2j+2k+2}\bigr) \cr
\noalign{\smallskip}
\bigl(x^{2k+1}\bigr) \cr
}\right]_{j=0,\ldots,n-1\atop k=0,\ldots,n\hfill} &(A.24)\cr
\noalign{\smallskip}
x^{2n+1} &\phantom{:}=\sum_{m=0}^n \widehat Q_{2m+1} \widehat b_{2m+1,2n+1} &(A.25)\cr
\noalign{\smallskip}
\widehat h_{2n+1} &\phantom{:}={Z_n^{({\rm o})}\over Z_{n-1}^{({\rm o})}} &(A.26)\cr
\noalign{\smallskip}
\widehat a_{2m+1,2n+1} &\phantom{:}={1\over Z_{n-1}^{({\rm o})}} \det\left[\matrix{
\bigl(c_{2j+2k+2}\bigr) \cr
\noalign{\smallskip}
\bigl(\delta_{m,k}\bigr) \cr
}\right]_{j=0,\ldots,n-1\atop k=0,\ldots,n\hfill} &(A.27)\cr
\widehat b_{2m+1,2n+1} &\phantom{:}={1\over Z_m^{({\rm o})}} \det\left[\matrix{
\bigl(c_{2j+2k+2}\bigr) \cr
\noalign{\smallskip}
\bigl(c_{2n+2k+2}\bigr) \cr
}\right]_{j=0,\ldots,m-1\atop k=0,\ldots,m\hfill}\,. &(A.28)\cr
}$$
These polynomials being unique, the relations above can also be proven by checking directly,
as for equation (2.6), that the polynomials (A.18) and (A.24) do satisfy the orthogonality condition
(A.2) with the normalization (A.3).

\medskip
\bigskip
\noindent{\elevenrmb Appendix B. Examples of block orthogonal polynomials for three subspaces}

\medskip
\noindent Let us only illustrate section 2.3 with the following polynomial examples for equation (2.29),
$N_1=N_2=1$, $N=3$, ${\cal E}_1={\cal P}_1$, ${\cal E}_{1,2}={\cal P}_2$ and ${\cal E}={\cal P}_3$,
with distinct inner products $(\,,\,)_{\rho,\sigma}$ defined by the positive measures
$\{{\cal D}_{\rho,\sigma},w_{\rho,\sigma}(x)dx\}$ for $1\le\rho<\sigma\le3$.
Then, each of the three one-dimensional vector spaces ${\cal E}_1$, ${\cal E}_2$ and ${\cal E}_3$ are defined,
respectively, by the monic basis polynomials,
$$\widehat P_0=1\qquad\widehat P_1=x+a_{0,1}\qquad\widehat P_2=x^2+a_{1,2}\,x+a_{0,2} \eqno(B.1)$$
where, $a_{0,1}$, $a_{1,2}$ and $a_{0,2}$ have to be determined from the orthogonality conditions,
$$(\widehat P_0\,,\,\widehat P_1)_{1,2}=0\qquad(\widehat P_0\,,\,\widehat P_2)_{1,3}=0
\qquad(\widehat P_1\,,\,\widehat P_2)_{2,3}=0\,. \eqno(B.2)$$
The first equation above defines uniquely ${\cal E}_2$ in all cases.
The last two equations may have for ${\cal E}_3$ a unique solution, no solution or an infinite number of solutions
depending linearly on one parameter.

\smallskip
\noindent (i) If, for $1\le\rho<\sigma\le3$, ${\cal D}_{\rho,\sigma}$ is symmetric with respect to
the origin and $w_{\rho,\sigma}$ is even, then one gets from parity arguments,
$$a_{0,1}=a_{1,2}=0\qquad a_{0,2}=-{c_{1,3;2}\over c_{1,3;0}} \eqno(B.3)$$
where, $c_{\rho,\sigma;n}:=\int_{{\cal D}_{\rho,\sigma}}dx\,w_{\rho,\sigma}(x) x^n$.
Thus, in that case, ${\cal E}_3$ exists and it is unique.

\smallskip
\noindent (ii) If,
$${\cal D}_{\rho,\sigma}:=[0,\infty)\qquad w_{\rho,\sigma}(x):=e^{-x} x^{z_{\rho,\sigma}-1}
\quad z_{\rho,\sigma}>0 \eqno(B.4)$$
one finds using the {\it Euler integral}
\note{See, e.g., [15] 6.1.1.},
$c_{\rho,\sigma;n}=\Gamma(z_{\rho,\sigma}+n)$, $a_{0,1}=-z_{1,2}$ and,
$$\pmatrix{
1 &z_{1,3}\cr
z_{2,3}+a_{0,1} &z_{2,3}\,(z_{2,3}+1+a_{0,1}) \cr
\noalign{\smallskip}
}\pmatrix{
a_{0,2} \cr
\noalign{\smallskip}
a_{1,2} \cr
}=-\pmatrix{
z_{1,3}\,(z_{1,3}+1) \cr
\noalign{\smallskip}
z_{2,3}\,(z_{2,3}+1)\,(z_{2,3}+2+a_{0,1}) \cr
}\,. \eqno(B.5)$$
In the generic case, e.g., $z_{1,2}=1$, $z_{2,3}=2$ and $z_{1,3}=3$, the determinant of the $2\times2$ matrix
above is nonzero and there is a unique solution for ${\cal E}_3$.
For, e.g., $z_{1,2}=1$, $z_{2,3}=2$ and $z_{1,3}=4$, the determinant vanishes while the rank of the augmented
matrix is $2$.
Then, there is no solution.
For the two linear equations (B.3) be the same, one must have
$$\left\{\matrix{
\hfill(z_{2,3}+a_{0,1})\,(z_{2,3}-z_{1,3})+z_{2,3} &= &0 \cr
\noalign{\smallskip}
(z_{2,3}+a_{0,1})\,(z_{2,3}-z_{1,3})+2z_{2,3}-z_{1,3}+1 &= &0 \cr
}\right.\ \Rightarrow\
\left\{\matrix{
z_{2,3}-z_{1,3}+1 &= &0\hfill \cr
\noalign{\smallskip}
\hfill a_{0,1} &= &0\,. \cr
}\right.
\eqno(B.6)$$
This case is excluded since $z_{1,2}=-a_{0,1}$ has to be positive.

\smallskip
\noindent (iii) In the same case as above, except for the inner product $(\,,\,)_{1,2}$ assumed to be defined
now by any domain ${\cal D}_{1,2}$ symmetric with respect to
the origin and any even weight function $w_{1,2}$, $a_{0,1}$ vanishes by parity and equations (B.3)
and (B.4) still hold.
Therefore, for $z_{2,3}-z_{1,3}+1=0$ with $z_{1,3}$ and $z_{2,3}$ positive, there is an infinite number of
subspaces ${\cal E}_3$ defined by the basis polynomial
$\widehat P_2=x^2-z_{1,3}\,(z_{1,3}+1)+a_{1,2}(x-z_{1,3})$
depending linearly on one arbitrary parameter $a_{1,2}$.

\medskip
\bigskip
\noindent {\elevenrmb References}

\bigskip
\item{[1]} Giraud B G, Mehta M L and Weiguny A 2004
Orthogonal polynomials sets with finite codimensions
{\it C. R. Phys.} {\bf 5} 781--7

\item{[2]} Giraud B G 2005
Constrained orthogonal polynomials
{\it J. Phys. A: Math. Gen.} {\bf 38} 7299--311

\item{[3]} Giraud B G, Weiguny A, Wilets L 2005
Coordinates, modes and maps for the density functional
{\it Nucl. Phys. A } {\bf 761} 22--40

\item{[4]} Hohenberg P and Kohn W 1964
Inhomogeneous electron gas
{\it Phys. Rev.} {\bf 136} B864--B871

\item{[5]} Dreizler R M and Gross E K U 1990
{\it Density Functional Theory}
(Berlin: Springer)

\item{[6]} Kohn W and Sham L J 1965
Self-consistent equations including exchange and correlation effects
{\it Phys. Rev.} {\bf 140} A1133--A1138

\item{[7]} Fiolhais C, Nogueira F and Marques M (ed) 2003
{\it A Primer in Density Functional Theory}
(Berlin: Springer)

\item{[8]} Messiah A 1999
{\it Quantum Mechanics}
(New York: Dover)

\item{[9]} Halmos P R 1958
{\it Finite-Dimensional Vector Spaces} 2nd edn
(Princeton: D Van Nostrand)

\item{[10]} Lang S 1965
{\it Algebra}
(Reading, MA: Addison-Wesley)

\item{[11]} Kreyszig E 1978
{\it Introductory Functional Analysis with Applications}
(New York: Wiley)

\item{[12]} Szeg\"o G 1975
{\it Orthogonal Polynomials} (Colloquium Publications vol XXIII)
4th edn
(New York: American Mathematical Society)

\item{[13]} Chihara T S 1978
{\it An Introduction to Orthogonal Polynomials}
(New York: Gordon and Breach)

\item{[14]} Bateman H 1953
{\it Higher Transcendental Functions} vol 2
(New York: McGraw-Hill)

\item{[15]} Abramowitz M and Stegun I A 1972
{\it Handbook of Mathematical Functions}
(New York: Dover Publications, Inc.)

\item{[16]} Gradshteyn I S and Ryzhik I M 2000
{\it Table of Integrals, Series, and Products} 6th edn
(New York: Academic)

\item{[17]} Bourbaki N 1970
{\it \'El\'ements de Math\'ematiques, Alg\`ebre}
(Paris: Hermann)
chapitres 1 \`a 3

\item{[18]} Mehta M L 1989
{\it Matrix Theory}
(Les Ulis, France: Les Editions de Physique)

\item{[19]} Dieudonn\'e J 1969
{\it Foundation of modern analysis}
(New York: Academic)

\item{[20]} Normand J-M 2004
Calculation of some determinants using the $s$-shifted factorial
{\it J. Phys. A: Math. Gen.} {\bf 37} 5737--62

\item{[21]} Aptekarev A J, Branquinho A and Van Assche W 2003
Multiple orthogonal polynomials for classical weights
{\it Trans. Amer. Math. Soc.} {\bf 355(10)} 3887--914

\item{[22]} Andrews G E, Askey R and Roy R 1999
{\it Special Functions (Encyclopedia of Mathematics and its Applications} vol 71{\it)}
(Cambridge: Cambridge University Press)

\end